\documentclass[letterpaper, 12pt]{article}
\pdfoutput = 1

\usepackage{shortcuts}

\usepackage[margin = 2.5cm]{geometry}
\titleformat{\subsubsection}
  {\normalfont\normalsize \it}{\thesubsubsection}{1em}{}

\numberwithin{equation}{section}

\usepackage[nottoc]{tocbibind}

\begin{document}

\thispagestyle{empty}

\begin{flushright}
\texttt{BRX-TH-6680}
\end{flushright}

\begin{center}

\vspace*{5em}

{\Large \textbf{The Ultraviolet Structure of Quantum Field Theories} \\ \medskip
  Part 1: Quantum Mechanics}

\vspace{1cm}

{\large \DJ or\dj e Radi\v cevi\'c}
\vspace{1em}

{\it Martin Fisher School of Physics\\ Brandeis University, Waltham, MA 02453, USA}\\ \medskip
\texttt{djordje@brandeis.edu}\\

\vspace{0.06\textheight}
\begin{abstract}
{This paper fires the opening salvo in the systematic construction of the lattice-continuum correspondence, a precise dictionary that describes the emergence of continuum quantum theories from finite, nonperturbatively defined models (``lattice theories''). Here the focus will be on quantum field theory in $(0+1)$D, i.e.\ quantum mechanics. The main conceptual achievement is an explicit and systematic procedure for reducing a theory with a large but finite Hilbert space to a subtheory in which wavefunctions satisfy prescribed smoothness and compactness constraints. This reduction, here named taming, in effect defines quantum mechanics on a continuum target space. When appropriate lattice theories are tamed, many familiar continuum notions explicitly emerge, e.g.\ canonical commutation relations, contact terms in correlation functions, continuous spacetime symmetries, and supersymmetry algebras. All of these are thus ``put on the lattice'' using the present framework. This analysis also leads to further insights into old subjects: for example, it is proven that any supersymmetric lattice theory must have a vanishing Witten index.}

\end{abstract}

\vfill
\textit{Dedicated to the victims of COVID-19.}
\end{center}

\newpage


\tableofcontents

\newpage

\section{Introductions}

\subsection{This series} \label{subsec intro ser}

This is the first paper in a series devoted to general aspects of the lattice-continuum correspondence in quantum field theory (QFT). The agenda of this series is to study a number of lattice theories (more precisely, quantum systems with Hilbert spaces of large but finite dimension) and to rigorously demonstrate how their subsystems, when judiciously chosen, exhibit emergent continuum behavior.

Understanding QFT from such a ``finitary'' point of view is immediately relevant to simulating strongly interacting quantum systems, either numerically or experimentally. Such an undertaking is also significant from a purely theoretical standpoint: it engenders a new perspective on the foundations of QFT, leads to rigorous nonperturbative definitions without invoking the often-forbidding machinery of functional analysis, and provides blueprints for constructing various new toy models. These formal matters will be the focus of this series.

To better understand the perspective offered here, it is helpful to briefly recall some fundamental ideas of modern QFT. Ever since lattice gauge theories afforded us a deep understanding of confinement and a springboard to numerically exploring hitherto intractable continuum field theories \cite{Wilson:1971bg}, a significant amount of work in high energy physics has been devoted to the interplay of lattice and continuum theories. Parallel developments in the condensed matter and statistical mechanics communities focused on using field-theoretic techniques to analyze specific lattice systems; see e.g.\ \cite{Cardy:1996xt, DiFrancesco:1997nk}. Over the years, these strands of research coalesced into a beautiful intellectual edifice. Its main tenet is that theories on very large lattices can be divided into universality classes, so that all theories in the same class have the same long-distance behavior: operator content, correlation functions, symmetries, spectral properties, and so on. These universal data can be extracted from individual lattice theories, often with the help of renormalization group techniques. Within each universality class, the universal data are then understood to be encoded by a \emph{continuum QFT} (cQFT). A cQFT is defined without explicit reference to any underlying lattice structure, as a collection of interdependent functions that assign numbers to manifolds, so that e.g.\ a partition function maps a spacetime to a real number, and a correlation function maps a collection of spacetime points to a complex number. Thus defined, cQFTs are useful because some of them obey stringent consistency conditions that allow us to infer a lot about them from relatively little input. Well known examples include theories with conformal symmetry, topological invariance, or supersymmetry. Within this Weltanschauung, particular lattice theories take a back seat: we focus on them only insofar as they allow us to extract universal data contained in some cQFT of interest.

The problem with this grand edifice is that its foundation is not solid. Understanding individual lattice theories is feasible only in some special cases. When interactions are strong on the lattice, it is usually impossible to find any universal data. In fact, generic lattice theories will not admit any continuum description. Conversely, understanding cQFTs without a lot of special structure is notoriously difficult, and relatively little can be done with a generic theory defined in an axiomatic framework \cite{Atiyah:1988, Haag:1996, Segal:1988}. In particular, it is often impossible to ``work backwards'' and identify a single lattice theory that could give rise to the desired universal data. In short, if the lattice-continuum correspondence is viewed as a map from lattice theories to cQFTs, the problem is that we understand neither the domain nor the range of this map. It should come as no surprise that we also understand little about how individual lattice states and operators map to continuum ones. This severely curtails our ability to understand --- or even contemplate! --- the microscopic details of our world.

Not all is doom and gloom, of course. Here are some general lessons we \emph{do} know about the lattice-continuum correspondence:
\begin{itemize}
  \item Lattice theories that are near a second-order phase transition (in parameter space) have continuum descriptions. At these points the correlation lengths of lattice operators are huge, and the corresponding correlation functions are insensitive to most small changes in the microscopics, like moving an operator insertion by a few lattice spacings. The corresponding cQFTs are conformal field theories and their deformations that slightly break the conformal invariance \cite{Pokrovsky:1968, Polyakov:1970xd}.
  \item Lattice theories with vanishing correlation lengths also admit continuum descriptions. The corresponding cQFTs are topological field theories or their cousins, symmetry-protected topological orders \cite{Chen:2011pg}.
  \item In general, a symmetry of a cQFT may be emergent, and the corresponding lattice theory need not exhibit it. However, if a cQFT has an \emph{anomalous} symmetry, any corresponding lattice theory must have this symmetry \cite{tHooft:1979rat}, and the symmetry must not be ``on-site'' \cite{Wen:2013oza}. A particularly famous incarnation of this phenomenon is the no-go theorem prohibiting lattice theories from having an on-site axial symmetry \cite{Nielsen:1980rz, Nielsen:1981xu}.
\end{itemize}

\textbf{The goal of this series of papers} is to extend these lessons by explicitly \hbox{\emph{constructing}} instances of the lattice-continuum correspondence. The principal strategy is to work with lattice theories \emph{without} focusing on correlation/partition functions as carriers of universal data. Instead, the focus will be on subsectors of lattice Hilbert spaces, or of lattice operator algebras, in which certain natural continuum rules hold up to an explicitly specified precision. \hbox{Trading} eminently physical workhorses (like correlation functions) for more abstract kinematic objects (like operator algebras) in order to track the emergence of continuum physics from a lattice will be the main conceptual novelty of these papers.

This approach will prove fruitful. This series will answer some old questions of the type ``how to put $X$ on a lattice?'', where $X$ includes chiral theories, supersymmetry, Chern-Simons theory, and more. Some of these questions have been the focus of lattice theorists for decades, the main goal being to numerically study phenomenon $X$. A \emph{caveat emptor} is in order: the answers provided here will be rather elegant, but the resulting lattice theories will not necessarily be \emph{efficiently} simulable.

It will also be pertinent to ask about the opposite direction: what interesting constraints on the space of cQFTs follow from requiring them to encode universal data of finite theories? The context here is that a cQFT typically depends on many parameters --- various couplings, background fields, the size and curvature of the underlying spacetime --- and so a given cQFT should be understood to calculate the universal data of a particular lattice theory only if its parameters satisfy certain constraints. This way the cQFT is viewed as an \emph{effective theory}, with parameter constraints that depend, at the very least, on the scale set by the lattice spacing.\footnote{For example, a cQFT may be used to calculate the universal part of the thermal free energy of a lattice theory, as long as the temperature is much smaller than the energy scale set by the lattice spacing.} Remarkably, novel constraints on cQFT parameters can be proven by demanding that the lattice-continuum correspondence hold. For instance, one basic lesson demonstrated here will be that, in \emph{any} spacetime dimension, a scalar cQFT that encodes data of a finite theory must be an effective QFT that features \emph{at least two distinct scales}.

The methods developed in this series extend those of two earlier papers that studied the lattice-continuum correspondence of fermionic theories in $(1+1)$D \cite{Radicevic:2019jfe, Radicevic:2019mle}. The crucial ingredient of that analysis was the fact that continuum operators must be \emph{smooth} functions of the spatial coordinates. In other words, a continuum regime can be identified as the subspace of a finite theory in which operators at nearby lattice points are \emph{constrained} to differ by small amounts. This constraint was straightforward to implement for pure fermions. In theories with bosons or more exotic particles, the construction is analogous but much more involved. This will be described in the present series, in multiple installments of increasing complexity.

Each part of this series will primarily focus on the lattice-continuum correspondence in a definite spacetime dimension. The emphasis will always be on constructing emergent continuum QFTs directly from lattice theories. This paper will lay the groundwork and construct models of quantum mechanics with continuous target spaces starting from models with discrete target spaces.\footnote{Lattice QFTs are usually allowed to have infinite-dimensional (continuous) target spaces, but in this series the term ``lattice theory'' will more narrowly refer only to theories with a finite-dimensional Hilbert space. In other words, both the real space and the target space must be discrete in a lattice theory. In this paper, the real space will consist of a single point.} The next installment will unite target space and position space continuity, leading to a construction of 2D scalar continuum theories \cite{Radicevic:2D}. Part 3 will generalize these constructions to 3D and will show how to include gauge fields \cite{Radicevic:3D}, and in Part 4 these lessons will be extended to 4D and nonabelian target spaces \cite{Radicevic:4D}.

\newpage

\subsection{This paper} \label{subsec intro pap}

Quantum field theory in $0+1$ spacetime dimensions is just ordinary quantum mechanics (QM). It is easy to visualize: simply picture a single quantum particle moving on a \emph{target space}, which can be either discrete (like vertices of a graph) or continuous (like a manifold). States labeled by different positions of the particle on the target are all mutually orthogonal.

QM is normally not associated with the kind of divergent behavior that is well known in higher-dimensional QFT. Nevertheless, its Hilbert space can be infinite-dimensional, and this is enough for subtleties to creep in. A simple example is supplied by the free particle moving on a real line, whose Hamiltonian is the unbounded operator
\bel{\label{def H free R}
  H = - \frac1{2m} \dder{}x.
}
This theory has no well defined energy eigenstates. Of course, all physicists are inured to the existence of nonnormalizable states, but let us reflect for a second: here is a theory with a well defined energy spectrum but without \emph{any} energy eigenstates. How is such a theory different from a theory with well defined energy eigenstates?

One way to answer this question is to hark back to the idea that continuum theories encode the universal data of lattice theories, as outlined in Subsection \ref{subsec intro ser}. In this paper, a \emph{lattice theory} is any QM theory whose target space is a finite set; conversely, a \emph{continuum theory} (cQM for short) has a manifold as the target space. If this manifold is noncompact, like $\R$, the states describable by this cQM all have continuous and normalizable wavefunctions. Roughly, they are universal --- insensitive to both short- and long-distance behavior.

With this distinction in mind, the free particle \eqref{def H free R} can be understood as a cQM that corresponds to a lattice theory in which \emph{none} of the energy eigenstates are universal. Indeed, if the particle on the line is viewed as an approximation of a particle freely hopping along a discrete array of $K \gg 1$ sites, there will exist no eigenstate which is supported on less than $O(K)$ sites. In other words, no eigenstate will be insensitive to the long-distance properties of the lattice. From this perspective, the fact that the free particle cQM has no energy eigenstates is actually unsurprising: a generic theory with a $K$-dimensional state space would also have no universal energy eigenstates. In fact, the free particle on $K$ sites is special because its low-energy spectrum \emph{is} universal, given by $E(p) = \frac{p^2}{2m}$ for $p \in \R$. A generic lattice theory would show even less structure than that.

The above discussion is a very, very rough sketch of the approach taken in this paper. Indeed, \textbf{the goal of the present paper} is to make such an analysis of universality totally precise. This will in turn lead to a constructive definition of continuum quantum mechanics that refers only to a finite theory.

Concretely, this paper will work with discretized versions of flat, one-dimensional target spaces. All lattice models presented here can be viewed as versions of a quantum particle moving on a ring of $K$ sites in the presence of some potential. Such systems are often called qudits or clock models. Among their possible Hamiltonians, particular emphasis will be given to those whose universal data are captured by four simple models of cQM: the free particles on a circle and an infinite line, the harmonic oscillator, and the supersymmetric harmonic oscillator. These are going to be the four \emph{universality classes} of interest.

For lattice theories in each of these four classes, this paper will identify subalgebras of operators that approximate the algebras of appropriate continuum theories. Any desired accuracy can be achieved by varying $K$ and parameters of the subalgebras. An important aspect of this subalgebra-identification procedure is that it is the \emph{same} for all lattice models. In fact, the procedure can be thought of as a generalization of the renormalization group (RG) in the following sense. The core idea of RG is to integrate out all degrees of freedom at high spatial momenta, which is equivalent to restricting to a subalgebra of operators that act only on low-momentum degrees of freedom. The decimations proposed here are also restrictions to subalgebras. However, instead of choosing them to contain operators of low spatial momentum (which is not a concept readily available in QM anyway), the subalgebras are defined to contain only operators that preserve certain desirable traits of wavefunctions.

Two wavefunction traits will be used: \emph{smoothness} and \emph{compactness}. Both are defined relative to a specific basis of states labeled by target space positions. Let $\qvec{\e^{\i\phi}}$ be these basis vectors, with $\phi = \frac{2\pi}K n$ for $n = 1, \ldots, K$. Then a state $\qvec{\psi}$ has a \emph{smooth} wavefunction $\greek y(\phi) \equiv \qprod{\e^{\i\phi}}{\psi}$ if
\bel{\label{intro smoothness}
  \left|\greek y\left(\phi + \tfrac{2\pi}K\right) - \greek y(\phi) \right| < \delta\_S
}
for some $1/K \ll \delta\_S \ll 1$. Further, a wavefunction will be called \emph{compact} if
\bel{\label{intro compactness}
  |\greek y(\phi)| = 0 \quad \trm{whenever} \quad \left|\phi - \phi\^{cl} \right| > \delta\_C,
}
for some fixed $\phi\^{cl}$ and $1/K \ll \delta\_C \ll 1$. (Note that $\greek y(\phi + 2\pi) = \greek y(\phi)$ holds at all times.) Smoothness can be interpreted as compactness in the natural Fourier space corresponding to the basis elements $\{\qvec{\e^{\i\phi}} \}$. A wavefunction that is both smooth and compact will be called \emph{tame}.\footnote{Another appropriate term for tame states is ``coherent,'' as their wavefunctions are wavepackets localized both in the target space and in the associated momentum space. However, this word may trigger unwanted associations with the process of decoherence, as well as with other constructions of coherent states in finite-dimensional Hilbert spaces, and so it will not be used here. The space of tame wavefunctions is a lattice version of a Schwartz space, i.e.~a space of smooth rapidly decreasing functions. The archetypical Schwartz function is a Gaussian, and indeed many tame wavefunctions encountered here will be lattice Gaussians.} Operators that preserve smooth or tame subspaces will form the Ersatz continuum algebras advertised above. Indeed, imposing the constraint \eqref{intro smoothness} on a lattice QM will be taken to \emph{define} the appropriate cQM.


The definition of cQM that was sketched above will be presented in detail in \textbf{Section \ref{sec clock}}. The working examples will be the universality classes of the free particle on the line and the circle. The main lesson is that the effective cQM of a particle on a one-dimensional target space must be defined using \emph{two} cutoffs, if the target is compact, or \emph{three} cutoffs if the target is not compact. Canonical position and momentum operators will then emerge from their discrete counterparts after a projection to the appropriate (smooth or tame) subalgebra.

Before proceeding with more involved examples, \textbf{Section \ref{sec notes}} will briefly describe how this construction of cQM touches on several other bodies of work. This will be a collection of basic ideas that may each develop into a separate research project over time.

In \textbf{Section \ref{sec sho}}, these simple-minded considerations will be applied to the nontrivial question of the emergence of the most famous interacting cQM, the simple harmonic oscillator (SHO), from a specific lattice model with a finite Hilbert space. This instructive exercise will precisely identify the subspace in which the continuum description holds. In addition, it will be shown that there exist simple lattice theories whose low-energy spectra include both continuum SHO eigenstates and nonsmooth wavefunctions that one might have na\"ively excluded from low-energy considerations.

Analogous issues in the path integral formulation will be discussed in \textbf{Section \ref{sec path int}}. Restricting to smooth wavefunctions in the canonical formalism translates to precise requirements on the jaggedness of trajectories that are summed over in the path integral. These constraints in turn determine when calculating path integrals can be reduced to doing Gaussian integrals. The cutoffs formulated in the previous Sections imply the existence of a critical temperature above which all control over the path integral is lost. This is the quantum-mechanical analog of a ``roughening transition'' that will be explored in more detail later in this series.

In Sections \ref{sec clock}--\ref{sec sho}, everything was happening on a single time slice. As path integrals come with an in-built lattice structure in the temporal direction, Section \ref{sec path int} will also define \emph{temporal smoothing}, which will be the key ingredient in actually calculating path integrals.

Prior to defining supersymmetric quantum theories, \textbf{Section \ref{sec ferm}} will discuss the QM of a single fermionic degree of freedom. This simple theory admits no continuum limit, as its Hilbert space is two-dimensional. Nevertheless, its path integral will prove to be a tractable and nontrivial example of temporal smoothing, which will here be shown to give rise to contact terms in cQFT correlation functions.

Finally, supersymmetric QM models will be studied in \textbf{Section \ref{sec SQM}}. The highlight of this Section is the proof of a version of the fermion doubling theorem: the Witten index \cite{Witten:1982df} of any finite supersymmetric theory \emph{must} be zero. As an example, natural lattice versions of the supersymmetric SHO will be shown to contain (at least) two copies of the continuum theory. More generally, this means that any quantum theory --- in particular, any QFT --- that is both supersymmetric and fully UV-finite must have at least two degenerate vacua.

\section{The clock algebra and its restrictions} \label{sec clock}

\subsection{Clock and shift operators} \label{subsec X Z}

The stage upon which this paper is set is a Hilbert space $\H$ whose dimension is a finite, but possibly very large, integer $K$. The elements of any orthonormal basis of this space can be visualized as $K$ points arranged in a circle. Fix one such basis, and let its elements be $\qvec{\e^{\i\phi}}$, where
\bel{
  \phi \equiv \frac{2\pi}K n \quad \trm{for}\quad 1 \leq n \leq K.
}

The set $\A$ of all operators acting on $\H$ is isomorphic to the algebra of complex matrices $\C^{K\times K}$. Schwinger \cite{Schwinger:1960} was the first to emphasize, if not to discover, that the algebra $\A$ is naturally generated by unitary operators $X$ and $Z$ that act as
\bel{\label{def Z X}
  Z\qvec{\e^{\i\phi}} = \e^{\i\phi} \qvec{\e^{\i\phi}}, \quad X \qvec{\e^{\i\phi}} = \qvec{\e^{\i(\phi - \d\phi)}},
}
where
\bel{
  \d\phi \equiv \frac{2\pi}K.
}
In other words, the set of all possible products of $X$ and $Z$ contains precisely $K^2$ operators that span the algebra $\A$. These two generators obey
\bel{\label{def Z X rels}
  X^K = Z^K = \1, \quad XZ = \e^{\i\, \d\phi} ZX.
}
Any operators satisfying these relations are said to form a \emph{clock algebra}. In accordance with their action \eqref{def Z X} in the chosen basis, $Z$ will be called a \emph{clock operator}, and $X$ a \emph{shift operator}. In the simple case $K = 2$, $X$ and $Z$ are the familiar Pauli matrices.

There is an obvious parallel between the clock/shift operators and the more familiar position/momentum operators. In fact, at least when $K \gg 1$, it is extremely tempting to define Hermitian operators $-\i \log Z$ and $-\i \log X$ as the canonical position and momentum variables found in a continuum theory of a quantum particle moving on a circle. Unfortunately, due to the finiteness of the Hilbert space, these operators do not obey the canonical commutation relations. (Indeed, there exist no matrices $A$ and $B$ that satisfy $[A, B] = \i \1$, as the trace of the l.h.s.~is always zero while the trace of the r.h.s.~is always nonzero.) More finesse is needed to move from the clock algebra to the continuum.

A useful example is the \emph{free clock model}, given by the Hamiltonian
\bel{\label{def H free clock}
  H = \frac{1}{2(\d\phi)^2} \left(2 - X - X\+\right).
}
The overall scaling does not influence the physics, but it will simplify future formul\ae. This theory describes a quantum particle freely hopping between neighboring sites of the $\Z_K$ target space. The Hamiltonian is diagonal in the eigenbasis of the shift operator $X$, whose eigenstates are
\bel{\label{def p}
  \qvec p \equiv \frac1{\sqrt K} \sum_{\phi\, =\, \d\phi}^{2\pi} \e^{\i \phi p} \qvec{\e^{\i\phi}}, \quad -\frac K2 \leq p < \frac K2.
}
The integer labels $p$ will simply be called \emph{momenta} in this paper, since there is no way to confuse them with spatial momenta found in higher dimensions. The clock and shift operators act on the shift eigenstates as
\bel{
  Z\qvec p = \qvec{p + 1}, \quad X \qvec p = \e^{\i p\, \d\phi} \qvec p,
}
with $\left\qvec{p = -\frac K2\right} \equiv \left\qvec{p = \frac K2\right}$. Their energies are
\bel{
  E_p = \frac1{(\d\phi)^2} (1 - \cos p\,\d\phi).
}
At $|p| \ll K$, the spectrum can be expressed as
\bel{
  E_p = \frac12p^2\left[1 + O\left(p^2/K^2 \right)\right],
}
and in particular it is independent of $K$ at leading order.

The disappearance of $K$ at low momenta suggests that this part of the spectrum is universal, i.e.~that it can be captured by a cQM. Indeed, up to irrelevant prefactors, the spectrum is the same as that of the free particle on a line \eqref{def H free R}, except here $p$ takes quantized values --- precisely what one expects in a continuum theory of a particle on a circle. However, the majority of states in the spectrum have $p \sim K$ and their energies are not universal.\footnote{An interesting almost-exception are states with $|p \pm K/2| \ll K$, whose energies depend on $K$ only through an additive constant. This means that these can also be captured by a cQM, but there is no cQM that can capture both edges of the spectrum, as they are separated by a nonuniversal rift.}

The energy eigenfunctions
\bel{\label{def psi p}
  \greek y_p(\phi) \equiv \frac1{\sqrt K} \e^{\i \phi p}, \quad \phi = n \d\phi, \quad 1 \leq n \leq K
}
are also universal at $|p| \ll K$. In this context, this just means that they can be rescaled and reinterpreted as slowly varying, square-integrable wavefunctions on the unit circle,
\bel{
  \greek y\^c_p(\phi) \equiv \frac1{\sqrt{\d\phi}}\greek y_p(\phi) =  \frac1{\sqrt {2\pi}} \e^{\i \phi p}, \quad \phi \in [0, 2\pi).
}

\subsection{The smooth subalgebra} \label{subsec smooth}

The notions of universality used in the previous Subsection were somewhat qualitative. In order to precisely talk about continuum theories and their wavefunctions, it is first necessary to decide on a set of axioms that cQMs have to satisfy. This is a rather delicate business when working directly in the continuum, as it requires introducing the theory of distributions, unbounded operators, von Neumann algebras, etc \cite{Gelfand:1964, Reed:1972}. The present approach will be significantly more elementary. Instead of being concerned with reproducing a set of formal continuum axioms, it is possible to start from the finite clock algebra \eqref{def Z X rels} and restrict it to a natural subalgebra that preserves the space of wavefunctions that smoothly vary along the clock positions. It will then turn out that all features of a cQM that may be relevant to a physicist are contained in this smoothness-preserving (or just ``smooth'') subalgebra. Indeed, as advertised in Subsection \ref{subsec intro pap}, this paper will take the stance that cQM theories can be defined via appropriate restrictions of clock algebras with $K \gg 1$. The comparison of this approach to more conventional ones will be discussed in Subsection \ref{subsec axioms}.

The key idea of this construction comes from the free clock model, in which (as just shown in Subsection \ref{subsec X Z}) a restriction to low momenta corresponds to a restriction to the universal part of the theory. With this in mind, define the \emph{smooth subspace}
\bel{
  \H\_S \equiv \trm{span}\{\qvec p\}_{-p\_S \leq p < p\_S},
}
for $1 \ll p\_S \ll K$. The basis vectors $\qvec p$ are given by the eigenstates \eqref{def p} of the free theory \eqref{def H free clock}, but this definition will be used to define smooth subspaces in a much wider class of interacting theories. The quantity $p\_S$ is the second of the three cutoffs that were advertised in the Introduction (the first one is, of course, $K$).

It is very convenient to work with smooth clock states
\bel{
  \qvec{\e^{\i\varphi}} \equiv \frac1{\sqrt{2p\_S}} \sum_{p = -p\_S}^{p\_S - 1} \e^{- \i p\varphi } \qvec p, \quad \varphi \equiv \frac{2\pi}{2p\_S} n, \quad 1\leq n \leq 2p\_S.
}
In terms of original clock eigenstates, smooth states are smeared over blocks of $\sim \frac K{2p\_S}$ sites,
\bel{\label{def varphi}
  \qvec{\e^{\i\varphi}} = \sum_{\phi \, = \, \d\phi}^{2\pi} f_{\varphi,\, \phi} \qvec{\e^{\i\phi}}, \qquad f_{\varphi, \, \phi} \equiv \frac1{\sqrt{2p\_S K}} \sum_{p = -p\_S}^{p\_S - 1} \e^{- \i p (\varphi - \phi)} = \frac{2\i}{\sqrt{2p\_S K}} \frac{\sin p\_S(\phi - \varphi)}{\e^{\i (\phi - \varphi)} - 1}.
}
It will also be convenient to define the smearing angle
\bel{\label{def d varphi}
  \d\varphi \equiv \frac{2\pi}{2p\_S} \gg \d\phi.
}

In terms of operators, defining an appropriate smooth algebra is slightly subtle. The na\"ive choice is to simply take all $2p\_S \times 2p\_S$ submatrices that act on $\H\_S$. However, this is not a unital subalgebra of the original algebra $\A$, i.e.~it does not contain the identity operator. Roughly speaking, no system described by a nonunital algebra can evolve unitarily, and so the smooth subsystem described by the na\"ive subalgebra would turn out pathological. To circumvent this issue, define the \emph{smooth subalgebra} $\A\^S$ to consist of all operators in $\A$ that have the form
\bel{\label{form smooth op}
  \bmat{\big(\ddots \big)_{\, 2p\_S \times 2p\_S}}00{\color{red} \trm{diag}\big(\ddots \big)_{(K - 2p\_S) \times (K - 2p\_S)}},
}
where the states \emph{outside} $\H\_S$ (red entries) are in the momentum basis. In other words, the basis elements of $\A\^S$ are obtained by projecting
\bel{\label{smoothing}
  X^n \mapsto X^n, \quad Z^p \mapsto (Z^p)\_S
}
in all the operators $X^n Z^p$, $1 \leq n, p \leq K$, that form the basis of the $K^2$-dimensional vector space over $\C$ that is the algebra $\A$. This projection, which will also be called \emph{smoothing}, leaves all powers of the shift operators $X$ invariant, but projects all powers of clock operators $Z$ to the space of operators of the form \eqref{form smooth op}. In particular, it is simple to check that $(Z^p)\_S = 0$ for $2p\_S \leq |p| \leq K - 2p\_S$. The resulting algebra is a vector space of complex dimension $(2p\_S)^2 + K - 2p\_S$.

The smooth algebra $\A\^S$ can be viewed as the set of all operators that act on
\bel{
  \widehat \H\_S \equiv \H\_S \oplus \left[ \bigoplus_{p = p\_S}^{-p\_S - 1} \trm{span}\{\qvec{p} \}\right].
}
All the Hilbert spaces in the bracket are one-dimensional. Somewhat nonstandardly, the direct sum is used to indicate the existence of different superselection sectors.\footnote{This smooth subalgebra is a slight refinement of certain types of subalgebras that were recently used to quantify target space entanglement in quantum mechanics \cite{Radicevic:2016kpf, Balasubramanian:2018axm, Mazenc:2019ety}. In particular, \cite{Mazenc:2019ety} considered subalgebras of operators of the  form \eqref{form smooth op} but with a one-dimensional center: any operator in their subalgebra was expressible as a direct sum of some $2p\_S \times 2p\_S$ matrix and a $(K-2p\_S) \times (K - 2p\_S)$ identity matrix.}

In general, smoothing does not commute with multiplication, i.e.\
\bel{
  (Z^{p_1 + p_2})\_S \neq (Z^{p_1})\_S (Z^{p_2})\_S,
}
though in many cases the two sides may agree. For instance, for any $p < K - 2p\_S$, one has $(Z^p)\_S = (Z\_S)^p$. On the other hand, for example, $\1 = \left(Z\+ Z\right)\_S \neq \left(Z\+\right)\_S Z\_S$. The important lesson is that there exist two different ways to define a product between two smooth operators, depending on whether one smoothes before or after multiplying the matrices.

The physical meaning of smoothing is evident in the clock basis. Any wavefunction of a state in $\H\_S$ can be written as a linear combination of momentum-basis wavefunctions \eqref{def psi p}
\bel{
  \greek y(\phi) = \sum_{p = -p\_S}^{p\_S - 1} \alpha_p \, \greek y_p(\phi) = \frac1{\sqrt K} \sum_{p = -p\_S}^{p\_S - 1} \alpha_p \, \e^{\i\phi p}, \quad \trm{with} \quad \sum_{p = -p\_S}^{p\_S - 1} |\alpha_p^2| = 1.
}
Such a wavefunction is smooth, in the precise sense that
\bel{\label{smoothness cond}
  \greek y(\phi + \d\phi) = \sum_{p = -p\_S}^{p\_S - 1} \alpha_p \, \e^{\i \phi p} \left(1 + \i p\, \d\phi + O\left(p^2\_S/K^2\right) \right) \approx \greek y(\phi) + \frac{\i\,\d\phi}{\sqrt K} \sum_{p = -p\_S}^{p\_S - 1} p\, \alpha_p \, \e^{\i\phi p}.
}
The ratio of the two momentum cutoffs, $p\_S/K$, controls the smoothness of wavefunctions. It is important to stress that a generic lattice QM has \emph{no} comparable requirement on wavefunctions: they are allowed to be arbitrarily jagged, even in the limit $K \gg 1$.

It is convenient to define a \emph{formal derivative} as a map that acts on an exponential function as
\bel{
  \hat\del_\phi \e^{\i\phi p} \equiv \i p \, \e^{\i\phi p}.
}
Then the smoothness condition \eqref{smoothness cond} can be written as
\bel{
  \greek y(\phi + \d\phi) \approx \greek y(\phi) + \hat\del_\phi \greek y(\phi) \, \d\phi.
}
In other words, restricting to $\H\_S$ can be understood as imposing a \emph{constraint}, or \emph{superselection rule}, that all wavefunctions have formal derivatives (which correspond to multiplication by $\i p$ in momentum space) equal to discrete derivatives (which are defined as differences in position space), at least to leading order in $p\_S/K$:
\bel{
  \del_\phi \greek y(\phi) \equiv \frac1{\d\phi}\left(\greek y(\phi + \d\phi) - \greek y(\phi) \right) \approx \hat\del_\phi \greek y(\phi).
}

Within the smooth subspace it also makes sense to focus on the \emph{momentum operator}
\bel{\label{def P}
  P \equiv \frac{X - X\+}{2\i \, \d\phi}.
}
(It is unchanged by smoothing, $P\_S = P$.) This operator acts on a smooth clock state as
\bel{
  P \qvec{\e^{\i\varphi}} =  \sum_{p = -p\_S}^{p\_S - 1} \frac{\e^{\i \varphi p}}{\sqrt{2p\_S}} \frac{\sin (p\,  \d\phi)}{\d\phi} \qvec p = \frac1{\sqrt{2p\_S}} \sum_{p = -p\_S}^{p\_S - 1} \! p \big(1 + O(\tfrac{p\_S}K )\big) \e^{\i \varphi p} \qvec p \approx -\i \hat \del_\varphi \qvec{\e^{\i\varphi}}.
}
It is thus \emph{only} within $\H\_S$ that $X$ can be thought of as an exponential of the derivative.

\subsection{The tame subalgebra} \label{subsec taming}

The previous Subsection considered a space of smooth wavefunctions on a circle, with an algebra of operators constructed in such a way that no operator can destroy the smoothness. It is also possible to start from a clock model and define the space of smooth wavefunctions with support on a line, i.e.\ on a set of clock positions that approximates the tangent space constructed at a particular point of the target. The projection to this subspace will be called \emph{taming}. It is defined as smoothing followed by a projection to a set of smooth states where $\varphi$ is restricted to an interval whose size is much smaller than the circle length ($2\pi$). The resulting space of states is the \emph{tame subspace} of the clock model,
\bel{
  \H\_T \equiv \trm{span}\left\{\qvec{\e^{\i\varphi}} \right\}\_{-\varphi\_T \leq \varphi < \varphi\_T}.
}
The idea is that this space contains only small and smooth fluctuations around $\varphi = 0$. The maximal size of these flucuations, $\varphi\_T$, is the third small parameter of a cQM. It is chosen to fit into the hierarchy of scales
\bel{\label{scale hierarchy}
  1 \gg \varphi\_T \gg \d\varphi \gg \d\phi.
}
It is also useful to define the ratio
\bel{\label{def nG}
  n\_T \equiv \frac{\varphi\_T}{\d\varphi}
}
so that $\dim\H\_T = 2n\_T$. The hierarchy of cutoffs \eqref{scale hierarchy} can then be expressed as the hierarchy of Hilbert space dimensions,
\bel{
  \dim \H = K \gg \dim \H\_S = 2p\_S \gg \dim \H\_T = 2n\_T \gg 1.
}

All tame states have low momenta, by construction. It is possible to define a set of states that are not necessarily smooth but in which the clock position $\phi$ is restricted to a certain interval. This would give rise to the subspace of \emph{compact} wavefunctions, defined as
\bel{\label{def HC}
  \H\_C \equiv \trm{span} \left\{\qvec{\e^{\i\phi}} \right\}\_{-\phi\_C \leq \phi < \phi\_C}.
}
This construction will not be of interest in this paper; all compact wavefunctions that will be considered will also be smooth, and hence they will be tame.

It is also possible to choose a subspace of small fluctuations around some nonzero position $\varphi\^{cl}$. Moreover, the smooth space $\H\_S$ can be reduced to a direct sum over many ($\sim p\_S/n\_T$) tame subspaces, each describing tame fluctuations around a different background $\varphi\^{cl}$. This will be the basis of algebraically defining certain solitonic field configurations in higher dimensions.
\newpage

The unital subalgebra of tameness-preserving operators $\A\^T$, or simply the \emph{tame subalgebra}, can again be defined to contain those operators that preserve the extension of the tame space $\H\_T$,
\bel{
  \widehat \H\_T \equiv \H\_T \oplus \left[ \bigoplus_{\varphi = \varphi\_T}^{-\varphi\_T - \d\varphi} \trm{span}\{\qvec{\e^{\i\varphi}} \}\right] \oplus \left[ \bigoplus_{p = p\_S}^{-p\_S - 1} \trm{span}\{\qvec{p} \}\right].
}
Now consider the projections from $\A\^S$ to $\A\^T$,
\bel{
  (X^n)\_S = X^n \mapsto (X^n)\_T, \quad (Z^p)\_S \mapsto (Z^p)\_T.
}
This taming removes all smooth elements of $X^n$ that are outside the $2n\_T \times 2n\_T$ block corresponding to states in $\H\_T$. Clock operators are also not spared: their taming is the projection to the $\H\_T$ block plus the $2p\_S - 2n\_T$ diagonal terms in the smooth subspace. For example, some explicit expressions for simple tame operators are
\bel{\label{ZG}
  Z\_T = \left[
          \begin{array}{*9{c}}
            \left[1 - \frac1{2p\_S}\right]\e^{-\i\varphi\_T} &  & \frac1{2p\_S} \e^{- \i\varphi\_T} &  &  &  &  &  &  \\
             & \hspace{-1em}\ddots &  &  &  &  &  &  &  \\
             \frac1{2p\_S} \e^{\i(\varphi\_T - \d\varphi)} &  & \hspace{-1em}\left[1 - \frac1{2p\_S}\right]\e^{\i(\varphi\_T - \d\varphi)} &  &  &  &  &  &  \\
             &  &  & \hspace{-1em}{\color{blue}\left[1 - \frac1{2p\_S}\right] \e^{\i\varphi\_T}} &  & \hspace{-1em}{\color{blue} 0} &  &  &  \\
             &  &  &  & \hspace{-1em} {\color{blue} \ddots}&  &  &  &  \\
             &  &  & \hspace{-4em}{\color{blue} 0} &  & \hspace{-4em}{\color{blue}\left[1 - \frac1{2p\_S}\right]\e^{-\i(\varphi\_T + \d\varphi)}} &  &  &  \\
             &  &  &  &  &  & \hspace{-1em}{\color{red} 0} &  &  \\
             &  &  &  &  &  &  & \hspace{-0.5em} {\color{red} \ddots } &  \\
             &  &  &  &  &  &  &  & {\color{red} 0} \\
          \end{array}
        \right],
}
\bel{\label{XG}
  X\_T \approx \left[
          \begin{array}{*{10}{c}}
            1 - \frac{\i\,\d\phi}{2} & - \frac{\d\phi}{\d\varphi} &  & - \frac{\d\phi}{(2n\_T - 1)\d\varphi} &  &  &  &  &  & \\
            \frac{\d\phi}{\d\varphi} & 1 - \frac{\i\,\d\phi}{2} &  &  &  &  &  &  &  & \\
            &  & \ddots &  &  &  &  &  &  &   \\
            \frac{\d\phi}{(2n\_T - 1)\d\varphi} &  &  & 1 - \frac{\i\,\d\phi}2 &  &  &  &  &  &  \\ &
            &  &  & {\color{blue}0} &  &  &  &  &  \\ &
            &  &  &  & \hspace{0em} {\color{blue} \ddots}&  &  &  &  \\ &
            &  &  &  &  & \hspace{0em}{\color{blue} 0} &  &  &  \\ &
            &  &  &  &  &  & {\color{red} \e^{\i p\_S \d\phi}} &  &  \\ &
            &  &  &  &  &  &  & {\color{red} \ddots } &  \\ &
            &  &  &  &  &  &  &  & {\color{red} \e^{-\i(p\_S + 1)\d\phi}}
          \end{array}
        \right].
}\newpage
As before, red matrix elements correspond to the nonsmooth states, with basis vectors chosen to be shift eigenstates $\qvec p$. Blue matrix elements correspond to the smooth but untamed states; the $2p\_S \times 2p\_S$ block of black and blue entries is in the basis of smooth clock states $\qvec{\e^{\i\varphi}}$. The signs of off-diagonal elements in the $2n\_T \times 2n\_T$ block alternate over both columns and rows, which is not apparent from the entries shown. The only Taylor expansion here is in $\d\varphi$ in the off-diagonal terms of $X\_T$. Note, finally,  that $Z\_S$ is nilpotent and hence not diagonalizable, and so it is impossible to define a taming that leaves $Z\_S$ diagonal the same way $X$ had remained invariant under smoothing.

Within the tame subspace it is natural to discuss both the momentum and position operators in their familiar cQM forms. The momentum operator $P$, defined in \eqref{def P}, after taming acts as
\bel{
  P\_T \qvec{\e^{\i\varphi}} \approx - \i\hat\del_\varphi \qvec{\e^{\i\varphi}}
}
only as long as $|\varphi| \ll \varphi\_T$. The reason is that the original operator $P$, which acts as a formal derivative on all smooth states, would take any smooth state out of the ``tangent space'' $-\varphi\_T \leq \varphi < \varphi\_T$ after sufficiently many applications --- something $P\_T$ is not allowed to do. Thus $P\_T$ must fail to act as a formal derivative in some states. It is also not difficult to use the explicit form \eqref{XG} to verify that the matrix $P\_T$ is in no way approximately equal to a discrete derivative operator when acting on the edges of the tame subspace.

The \emph{position operator}
\bel{\label{def Q}
  Q \equiv \frac{Z - Z\+}{2\i}
}
after taming acts as
\bel{
  Q\_T \qvec{\e^{\i\varphi}} \approx \varphi \qvec{\e^{\i\varphi}}
}
on all states in $\H\_T$. More generally, this operator acts on the smooth subspace $\H\_S$ as
\bel{
  Q\_T \qvec{\e^{\i\varphi}} \approx \sin\varphi \qvec{\e^{\i\varphi}}.
}
When computing the higher powers $(Q^p)\_T$, the above approximate expressions are insufficient; off-diagonal elements will become relevant as $p$ is increased.

The commutator of the tame position and momentum operators, $[Q\_T, P\_T]$, is a matrix whose upper $2n\_T \times 2n\_T$ block has $\pm \i$ on \emph{all} off-diagonal entries, with zeroes on the diagonal. This is a far cry from the canonical commutation relation of the form $[q, p] = \i \1$. However,
\bel{\label{QP commutator}
  [Q, P]\_T = \left(\frac\i4 + O(\d\phi)\right) \left( (Z + Z\+) (X + X\+) \right)\_T
}
satisfies $[Q, P]\_T \qvec{\e^{\i\varphi}} \approx \i\, \cos\varphi \qvec{\e^{\i\varphi}} \approx \i \qvec{\e^{\i\varphi}}$ when acting on $\qvec{\e^{\i\varphi}} \in \H\_T$. This \emph{is} the canonical commutator. To get this result, it is imperative to multiply the operators \emph{before} taming.

\newpage

\section{Brief remarks on smoothing and taming} \label{sec notes}

\subsection{Renormalization and chaos in QM}

The coarse-grainings described so far can be understood as an extension of Wilsonian RG to QM. This extension is not unique. The issue is that the Wilsonian treatment is naturally defined in momentum space of a QFT, where degrees of freedom living in Hilbert spaces at high momenta can be integrated out to give effective low-momentum theories. In QM, there is no spatial momentum, and an analogous coarse-graining of the target space is impossible because, generically, the target space does not decompose into a direct product over momentum modes. In other words, a QM Hilbert space does not have any Fock structure. Integrating out high-momentum degrees of freedom is not naturally defined in QM, and so an alternative must be found. Section \ref{sec clock} presented one such alternative.

There exist other prescriptions for extending RG ideas to QM, e.g.\ \cite{Gupta:1993id, Polonyi:1994pn}. In particular, the algebraic approach presented here can be contrasted with earlier attempts in the same direction \cite{Radicevic:2016kpf, Ho:2017nyc}, in which the idea was to decimate all the way down to an Abelian subalgebra of operators and use the resulting density matrices as probes of spectral universality. That philosophy is orthogonal to the current one. Here, the focus is on coarse-grainings that give unitary, well defined quantum theories that exhibit continuum traits, while in the older papers, the resulting theories were decidedly not unitary (in fact, a theory described by an Abelian algebra is simply a theory of a classical probability distribution evolving in time).

This dichotomy highlights the fact that spectral universality, the mainstay of quantum chaos, is \emph{not} captured in full by continuum tools available in a clock model. To study quantum chaos, it is necessary to coarse-grain down to a classical theory with a $K$-dimensional state space and a $K$-dimensional operator algebra. To study the continuum limit of a lattice theory, on the other hand, one focuses on a quantum theory with a $\sim p\_S$-dimensional Hilbert space and $\sim p\_S^2$ independent operators. At $p\_S \ll K$, the latter space is much smaller than the former one, and consequently the continuum theory cannot be expected to capture all aspects of spectral statistics. (This argument does not apply to theories of fermions, where continuum descriptions are not reached by smoothing as described here; see Section \ref{sec ferm}.)

One facet of this observation is that the signatures of chaos captured by cQM (or cQFT) probes are primarily perturbative ones, such as Lyapunov exponents describing the short-time evolution of certain out-of-time-order correlation functions \cite{Shenker:2013pqa, Kitaev:2015}. In modern parlance \cite{Cotler:2016fpe}, the continuum knows about the slope but not the ramp or the plateau in spectral form factors. A provocative consequence of this line of thought is that any discussion of whether, say, $\mathcal N = 4$ super-Yang-Mills captures the plateau with or without disorder averaging may be moot unless a completely finite lattice definition of the $\mathcal N = 4$ theory is specified first.
\newpage

\subsection{Flowing to the continuum}

Smoothing and taming are two types of coarse-graining that can be defined in any clock model. Like all algebraic decimations, they induce a map between density matrices: any density matrix $\varrho$ can be mapped to a reduced density matrix $\varrho'$  by projecting onto the chosen subalgebra. 
If $\varrho$ is a thermal density matrix, then its reduction to $\varrho'$ defines the effective Hamiltonian $H'$ that governs the unitary part of the evolution of degrees of freedom described by the subalgebra. For more on this, see \cite{Lin:2018bud}.

A special situation arises if there exists a set of energy eigenstates that all remain pure after a reduction. Time evolution under the original Hamiltonian does not destroy the purity of any state in this subspace. The pure subspace and its effective Hamiltonian thus form a bona fide quantum theory on their own. The coarse-graining can then be said to cause the original theory to \emph{flow} to this new one. In most natural examples, this invariant subspace lies at the edge of the spectrum, and so it is common to refer to the corresponding subtheory as a ``low-energy theory.'' When expressed in a natural position space, however, the effective Hamiltonian of this subtheory may look very different from the original one. This paragraph is, of course, simply a pr\'ecis of the Wilsonian renormalization group (RG) ideas, formulated slightly more abstractly than is usual.

Flowing, in the narrow sense defined above, is different from merely restricting to a continuum theory, as described in Section \ref{sec clock}. In fact, the notion of a flow introduces a useful distinction between the kinds of cQMs one can get by restricting to the Ersatz continuum subalgebras $\A\^{S/T}$. If smoothing or taming \emph{do} cause a clock model to flow to a new quantum theory, then the cQM has universal eigenstates. If there is no energy subspace invariant under the desired decimation, i.e.\ if there is \emph{no} flow to a properly defined theory, the cQM still exists, but it must be like \eqref{def H free R}: an incomplete theory without well defined eigenstates.

The free clock model \eqref{def H free clock} illustrates these ideas well. Smoothing preserves a $2p\_S$-di\-men\-sio\-nal energy eigenspace --- all of $\H\_S$. In the natural position basis, given by the smooth clock states $\{\qvec{\e^{\i\varphi}}\}$, the effective Hamiltonian acts as
\bel{
  H'\qvec{\e^{\i\varphi}} \approx -\frac12 \hat\del_\varphi^2 \qvec{\e^{\i\varphi}},
}
with corrections suppressed by $O(p\_S^2/K^2)$. This is the Hamiltonian of a free particle in the continuum, cf.\ \eqref{def H free R}, but with the condition $\varphi \equiv \varphi + 2\pi$. On the other hand, taming does not preserve the purity of any energy eigenstate. This means that the free clock model flows to a cQM on a circle, but \emph{not} to a cQM on a line; it is \emph{impossible} to restrict to small fluctuations while keeping the quantum theory fully defined. Section \ref{sec sho} will exhibit a clock model which flows to a continuum theory under taming.

\subsection{Comparison to axiomatic approaches} \label{subsec axioms}

As mentioned at the beginning of Subsection \ref{subsec smooth}, there already exists an extensive operator-algebraic machinery that provides a rigorous foundation for cQM without invoking a lattice. How does the present approach compare?

A blitz summary of the conventional axiomatics could go as follows \cite{Gelfand:1964}. Given a target manifold $\Tbb$ (typically $\R$ or $S^1$), first consider the space $L^2(\Tbb)$ of square-integrable functions on $\Tbb$. Next, pick an operator $\Delta$ on this space (typically a differential operator like $H$ in \eqref{def H free R}). Then, define the domain of $\Delta$ as the set $\trm{Dom}(\Delta) \subseteq L^2(\Tbb)$ on which $\Delta$ is well defined (e.g.\ a set of smooth or differentiable functions on $\Tbb$). Finally, define the set $\trm{Dom}^\times(\Delta)$ of complex functions on $\Tbb$ that are not necessarily in $L^2(\Tbb)$ but that may be integrated against any element of $\trm{Dom}(\Delta)$ to give a finite answer. 
Instead of talking of a single Hilbert space, a cQM is then defined based on the so-called \emph{rigged Hilbert space}
\bel{
  \trm{Dom}(\Delta) \subseteq L^2(\Tbb) \subseteq \trm{Dom}^\times(\Delta).
}
The ``physical'' wavefunctions --- i.e.\ the states of the greatest interest to physicists --- are contained in $\trm{Dom}(\Delta)$. The other two spaces are needed to define the ``unphysical'' but logically and computationally necessary wavefunctions, such as the nonnormalizable eigenfunctions $\e^{\i p x}$ of the derivative operator on $\R$. It is the interplay between these three spaces, and the operator algebras built upon them, that is the foundation of rigorous cQM.

It is instructive to compare the rigged Hilbert space to one sequence of lattice Hilbert spaces defined in this paper,
\bel{
  \H\_T \subseteq \H\_C \subseteq \H,
}
where $\H\_T$ is the space of tame states, $\H\_C$ the space of compact states defined in \eqref{def HC}, and $\H$ the space of all possible states on the lattice target space. Taking this lattice to be a fine discretization of the manifold $\Tbb$ reveals both similarities and differences between this structure and the rigged Hilbert space:
\begin{enumerate}
  \item The defining feature of both $L^2(\Tbb)$ and $\H\_C$ is that their wavefunctions decay fast enough outside of some subset of $\Tbb$.
  \item Both $\trm{Dom}(\Delta)$ and $\H\_T$ are defined as spaces whose elements vary sufficiently slowly.
  \item Unlike $\H$, $\trm{Dom}^\times(\Delta)$ does not contain every imaginable function from $\Tbb$ to $\C$.
  \item $\trm{Dom}(\Delta)$ and $\trm{Dom}^\times(\Delta)$ depend on an operator $\Delta$ whose definition may be subtle, especially if it includes a singular potential. The definition of $\H\_T$ and $\H\_C$ in terms of $\H$ only depends on two integers.
\end{enumerate}
\newpage

In short, the definition of a cQM proffered in Section \ref{sec clock} is not a facsimile of the standard one, but it does exhibit significant structural similarities. The main conceptual difference is that the starting point in the present definition is manifestly (hyper)finite, and therefore easy to define. At no point is it necessary to prove the existence of the space $\H$ or its subspaces, and all the details that were omitted in the above summary of cQM axioms --- say, the choice of the topology or measure on $\Tbb$ that goes into the definition of $L^2(\Tbb)$, or the precise definition of the operator $\Delta$ --- are either not needed or are naturally induced by simple smoothing/taming procedures from the space $\H$. Perhaps most importantly, from a physicist's point of view, the definition of cQM via smoothing never assumes that there exist ``unphysical'' wavefunctions: all wavefunctions are on the same footing within the space $\H$, and states that become mixed upon coarse-graining are no more or less physically meaningful than those that remain pure.

\subsection{The quantum phase problem}

Issues surrounding the fact that no matrices can satisfy $[Q, P] = \i \1$ have historically fallen under the header of ``the quantum phase problem,'' whose roots reach back to Dirac's seminal paper on the quantization of electromagnetic fields \cite{Dirac:1927dy}. For more details, see \cite{Carruthers:1968my, Lynch:1995}. The crux of the problem, as pointed out by Susskind and Glogower \cite{Susskind:1964zz}, is that ladder operators $a$ and $a\+$ of the harmonic oscillator --- i.e.~operators that satisfy $[a, a\+] = \1$ --- \emph{cannot} be written in terms of two Hermitian operators, the number $n$ and the phase $\theta$, via
\bel{\label{polar decomp}
  a = n^{1/2} \e^{\i \theta}, \quad a\+ = n^{1/2} \e^{-\i\theta}.
}
In other words, there is tension between a quantum boson's canonical commutation relations and the fact that both $n$ and $\theta$ are observables in classical electromagnetism. Much work has been done to find acceptable phase operators in terms of the canonical operators $a$ and $a\+$.

This paper offers a very simple resolution of the quantum phase problem: since there exists no finite quantum theory with canonical commutation relations, the requirement $[a, a\+] = \1$ should be substituted by a much weaker one. Concretely, take a clock model and let
\bel{
  a \equiv \frac{Q + \i P}{\sqrt 2}, \quad a\+ \equiv \frac{Q - \i P}{\sqrt 2},
}
where $Q$ and $P$ are position and momentum operators from \eqref{def Q} and \eqref{def P}. These are well defined operators that always admit the polar decomposition \eqref{polar decomp}, and after taming they obey $[a, a\+]\_T \qvec{\e^{\i\varphi}} \approx \qvec{\e^{\i\varphi}}$. This highlights a deeper lesson: classical mechanics \emph{must} be understood as a particular limit of a cQM, and all lore about the correspondence principle or the quantization of classical theories only makes sense in a continuum context.

\subsection{Signatures of finiteness in the continuum}

The cQMs constructed via the coarse-grainings of Section \ref{sec clock} are effective theories, in the sense described in the Introduction: they come with in-built scales set by the large integers $K$, $p\_S$, and $n\_T$, and these scales set limits on the regime in which the continuum description is valid. It is standard lore that an effective theory receives corrections near the boundary of its domain of validity, regardless of whether the corrections come from a lattice theory or from a continuum theory that RG-flows to the effective theory at hand. These corrections are most obviously present when studying physics at high energies or temperatures, but they are also detected by e.g.\ the long-time behavior of the spectral form factor \cite{Cotler:2016fpe}, or by correlators of a large number of operators \cite{Ghosh:2017pel}.

Within QM, the implications of a UV cutoff --- often styled as a ``minimum length scale'' --- have long been an object of study. (See e.g.\ \cite{Kempf:1994su, Chang:2011jj} and references therein.) A ubiquitous focal point of this body of work has been the analysis of UV corrections to the Heisenberg uncertainty relations, or (conversely) of minimal lengths that are implied by modifying these relations. This paper provides a few instructive lessons concerning these questions.

First, a trivial observation: if a cQM is defined by merely smoothing a lattice theory, talking about canonical commutation relations (and hence about uncertainty relations) will not be meaningful. To do so, it is necessary to tame, not just smoothe.

A more substantial statement is that in a cQM defined by taming a lattice theory, there will exist \emph{three} competing corrections to the canonical commutation relations. The commutator $[Q, P]\_T$, calculated in eq.\ \eqref{QP commutator}, technically takes the form
\bel{
  [Q, P]\_T = \i \Big( 1 + O(\varphi\_T^2) + O(\d\varphi) + O\left((\d\phi)^2/(\d\varphi)^2\right) \Big).
}
The three correction terms can a priori have either sign, and therefore the task of proving bounds on the absolute value of this commutator is not quite trivial.

Since $\varphi\_T \gg \d\varphi$, it is somewhat natural to also assume $\varphi\_T^2 \gg \d\varphi$ and to drop the $O(\d\varphi)$ term (though note that this condition does not follow from the hierarchy \eqref{scale hierarchy}). The remaining two terms are comparable if
\bel{
  \varphi\_T \sim \frac{\d\phi}{\d\varphi}, \quad \trm{or} \quad p\_S^2 \sim n\_T K.
}
In the following Section it will be argued that this is, in fact, the natural parameter regime with which to define the harmonic oscillator cQM. Thus, in principle, these terms should both be kept. It appears that only the $O\left((\d\phi)^2/(\d\varphi)^2\right)$ term has been studied in the literature so far (see \cite{Jizba:2009qf} for a clear exposition, but beware: that reference takes $\d\varphi \sim \d\phi$, in the current notation, as one of its cases). There is clearly room for further results in this direction.

\newpage

\section{The simple harmonic oscillator} \label{sec sho}

Section \ref{sec clock} was concerned with a free theory in which smoothing corresponded to a clear-cut projection onto an energy eigenspace. This Section will study an interacting clock model in which this simplification does not occur. Remarkably, the smoothing procedure is still meaningful: the low-energy eigenstates are smooth (in fact, tame), and hence the model of this Section provides a nontrivial example of a lattice system that flows to a cQM theory of a particle in a line in the presence of a quadratic potential. This cQM is, of course, the ubiquitous simple harmonic oscillator (SHO).

Consider the clock model governed by the Hamiltonian
\bel{\label{def H sho clock}
  H_g = \frac{g^2}{2(\d\phi)^2} \left(2 - X - X\+ \right) + \frac1{2g^2} \left( 2 - Z - Z\+ \right).
}
The single coupling that controls the dynamics of this system is
\bel{
  \gamma \equiv \frac g{\sqrt{\d\phi}}.
}
At $\gamma \gg 1$ and $\gamma \ll 1$ the theory becomes the free clock model \eqref{def H free clock}. Nontrivial behavior happens when $\gamma$ is between these extremes. In fact, the theory (after a rescaling of $H_g$) enjoys a strong-/weak-coupling duality that interchanges shift and clock operators, and the self-dual point is precisely $\gamma = 1$. This is a caricature of a ``critical point,'' and here the theory comes closest to a continuum SHO.

\begin{figure}[t!]
  \centering
  \includegraphics[width = 0.31\textwidth]{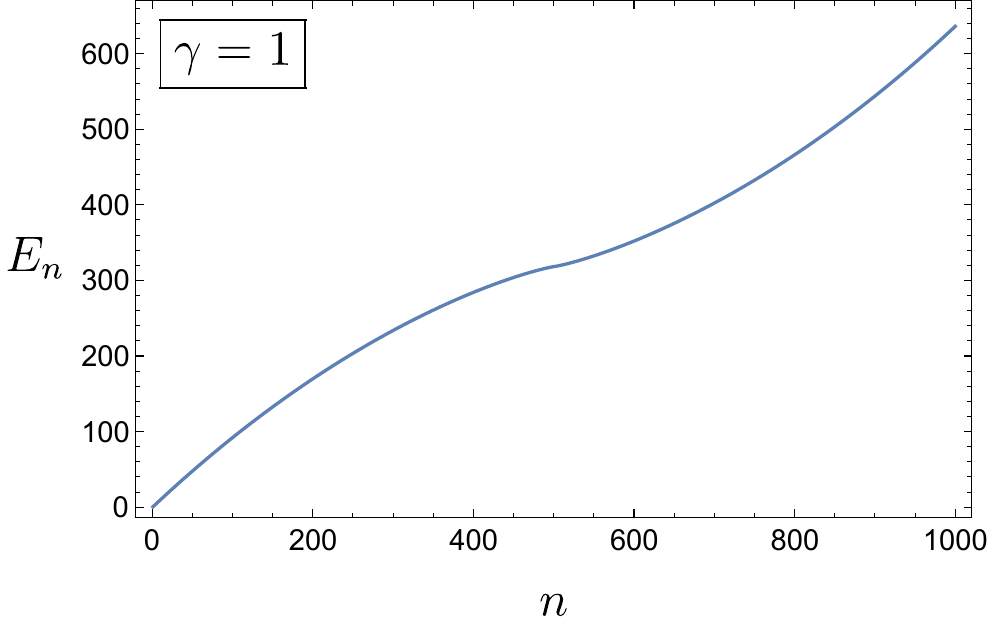}\quad
  \includegraphics[width = 0.31\textwidth]{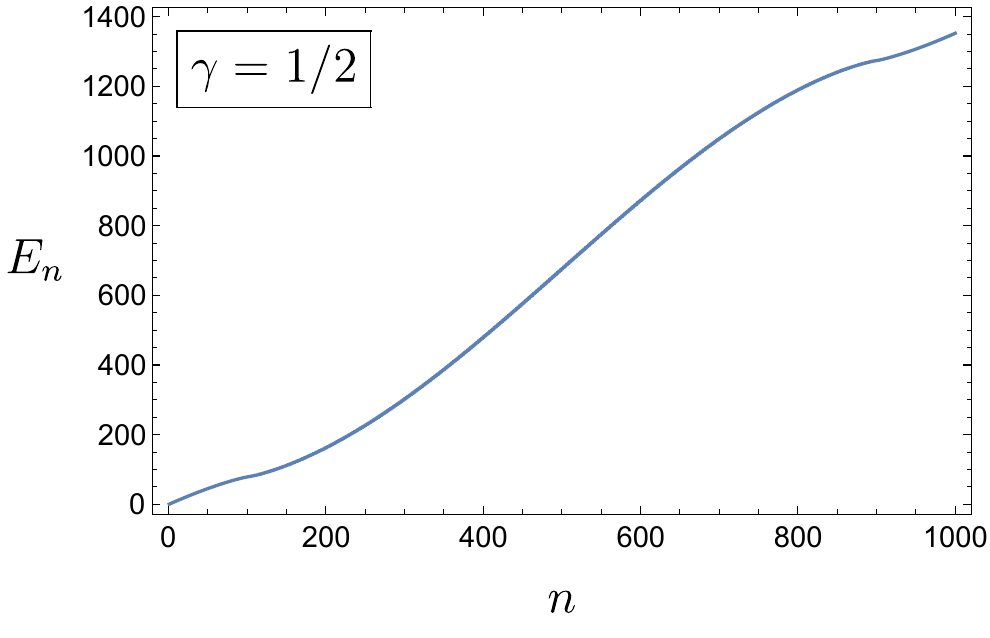}\quad
  \includegraphics[width = 0.31\textwidth]{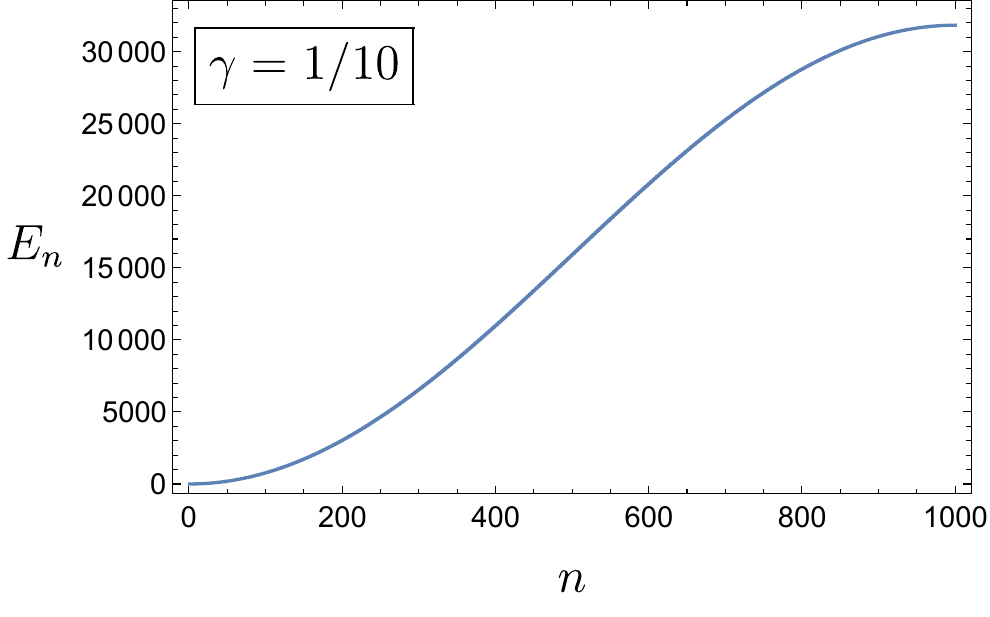} \\
  \includegraphics[width = 0.4\textwidth]{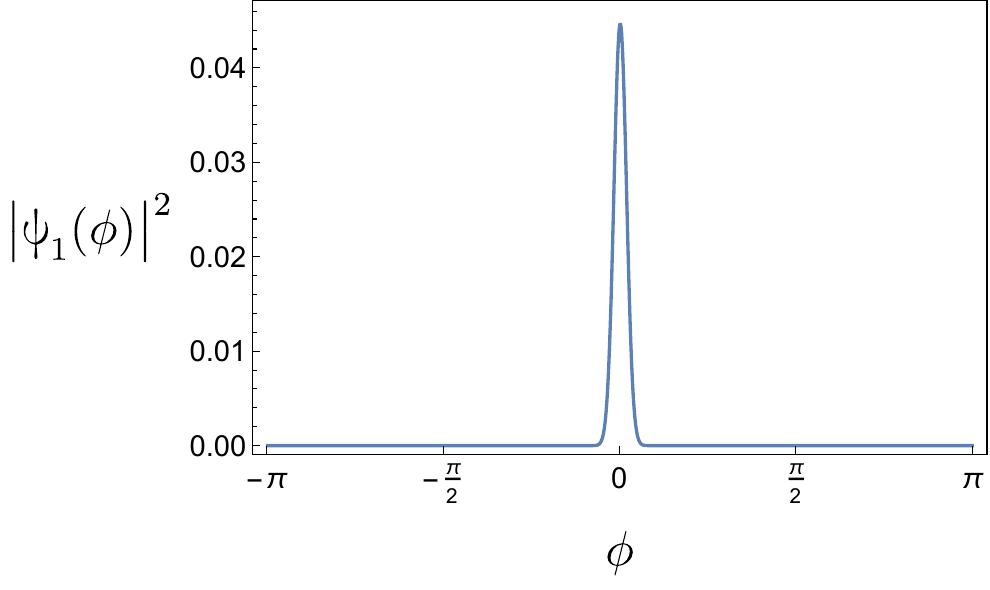}\qquad
  \includegraphics[width = 0.4\textwidth]{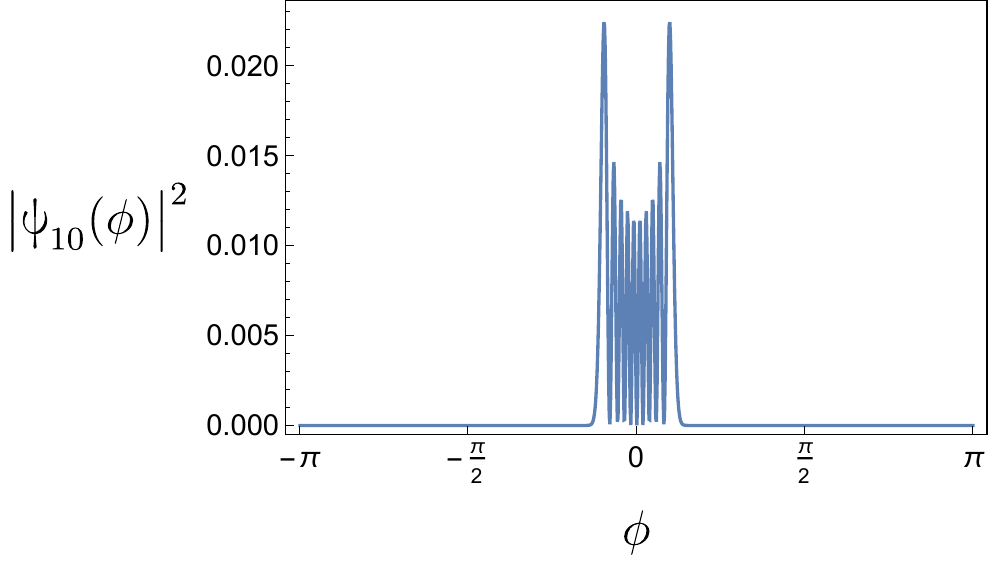}
  \caption{\small \emph{Top row}: The spectrum of the Hamiltonian $H_g$ in \eqref{def H sho clock} at various values of the coupling $\gamma$, for $K = 1000$, with $n = 1, 2, \ldots, K$ labeling the energies $E_n$ in increasing order. The shape of the spectrum is invariant under $\gamma \mapsto 1/\gamma$; the overall scale is irrelevant. At the self-dual point $\gamma = 1$, the low-energy states organize into a linear spectrum, as do the high-energy ones. This property disappears as $\gamma$ is changed, and already at $\gamma = 1/10$ the low-energy spectrum follows the quadratic dependence characteristic of the free clock model \eqref{def H free clock}. \emph{Bottom row}: two representative low-energy eigenfunctions at $\gamma = 1$, corresponding to the first and tenth lowest eigenstate. These eigenfunctions are good approximations to the eigenfunctions of the SHO cQM.} \label{fig sho spectra}
\end{figure}

Looking at the energy spectrum is a quick and dirty way to identify a set of candidate SHO states. The theory is straightforward to numerically diagonalize. The spectrum is shown on Fig.\ \ref{fig sho spectra}. At a fixed $K$, the share of states with a linear spectrum is maximized at $\gamma = 1$. The wavefunctions $\greek y_n(\phi)$ of low-energy states in the linear part of the spectrum are tame, i.e.\ localized in some window around $\phi = 0$ (in position space) and around $p = 0$ (in momentum space). They oscillate within this window with a characteristic wavelength that is always much greater than the ``lattice spacing'' $\d\phi$. This is a QM analogue of Wilsonian universality: as the coupling is tuned to a ``critical'' point, the low-energy states become insensitive to most microscopic details.

For each localized low-energy state, there exists a localized high-energy state centered around $\phi = \pi$ and whose wavefunction fluctuates with microscopic wavelengths $\sim \d\phi$. (The latter feature will be called \emph{a $(-1)^\phi$ modulation}.) The spectrum is linear there, too. These eigenspaces are related by the map $(Z, X) \mapsto - (Z,X)$, which is not a symmetry but merely flips the sign of the Hamiltonian and shifts it by an irrelevant constant. None of these high-energy states are tame because the $(-1)^\phi$ modulation makes them nonsmooth.

To precisely define tame states it is necessary to specify two parameters, $p\_S$ and $n\_T$.\footnote{It is often more intuitive to talk about eigenfunctions $\greek y(\phi)$ and to describe their tameness in terms of the smearing scale $\d\varphi$  and the position cutoff $\varphi\_T$, both expressed in units of the ``lattice spacing'' $\d\phi$. This amounts to talking about the small ratios $p\_S/K$ and $n\_T/K$ instead of the large numbers $p\_S$ and $n\_T$ themselves. This paper will use both points of view at different times.} In the free clock model, the Hamiltonian had no bearing on the choice of the smoothness parameter $p\_S$; for any choice, all energy eigenstates at momenta $|p| < p\_S$ would remain pure after smoothing. Conversely, for any $n\_T$, even the lowest eigenstate would become significantly mixed after taming. This was another special property of the free theory. The spectrum of $H_g$, shown on Fig.\ \ref{fig sho spectra}, suggests that even at the special point $\gamma = 1$ most states cannot be tame no matter how large $p\_S$ and $n\_T$ are. This means that the the microscopic theory \emph{dynamically induces} values, or at least bounds, for both $p\_S$ and $n\_T$.

To see a simple example of bounds imposed by the dynamics, fix $\gamma = 1$ and $K = 1000$. The ground state wavefunction $\greek y_1(\phi)$, whose modulus squared is depicted on the bottom left of Fig.\ \ref{fig sho spectra}, fits to a Gaussian with standard deviation $\sigma \approx 0.079$. In lattice parlance, this wavefunction is smeared over $2\sigma/\d\phi \approx 25$ sites; this number grows with $K$ at fixed $\gamma$. Excited states $\greek y_n(\phi)$ feature $n - 1$ nodes separated by $\Delta_n \phi \equiv 2\sigma/f(n)$, where $f(n) \geq 1$ grows slowly with $n$.

This observation implies that if $\varphi\_T$ is taken to be smaller than $\sigma$, no eigenfunction $\greek y_n(\phi)$ will remain pure after taming. This gives the most rudimentary bound on $\varphi\_T$ that follows from the dynamics of $H_g$. Furthermore, the rough estimate for the distance $\Delta_n \phi$ between nodes implies that, for $\varphi\_T > \sigma$, at least the first $[\varphi\_T/\sigma]$ states in the spectrum will remain pure upon taming. In other words, the dimension of the tame subspace $\H\_T$ will satisfy $2n\_T \geq [\varphi\_T/\sigma]$. By eq.\ \eqref{def nG}, this translates to saying that the smoothness cutoff must be chosen so that $\d\varphi \leq 2\sigma$, or so that $p\_S \geq [\pi/2\sigma] \approx 20$.

The bounds $\varphi\_T \geq \sigma$ and $\d\varphi \leq 2\sigma$ are very weak, but they illustrate the point: an interacting theory induces bounds on possible taming parameters. More interesting, physically, are the \emph{upper} bounds on $n\_T$ and $p\_S$ that may follow from the fact that the majority of states are not tame. In order to study these, it is helpful to notice that at $\gamma = 1$ the self-duality implies that the wavefunctions must take the same form in position space and in momentum space. In other words, for any energy eigenstate $\qvec n$ it is true that
\bel{
  \greek y_n(\phi) = \qprod p n,
}
where $\qvec p$ is a shift eigenstate with momentum $p = \phi/\d\phi$. This means that smoothness and compactness of these wavefunctions are related. If $\greek y_n(\phi)$ is compact, i.e.~if it is localized in a region $-\varphi\_T \leq \phi < \varphi\_T$, then in momentum space this wavefunction must be localized in the region $-\frac{\varphi\_T}{\d\phi} \leq p < \frac{\varphi\_T}{\d\phi}$. It is thus natural to choose
\bel{\label{relation pS nG}
  p\_S = \frac{\varphi\_T}{\d\phi} = n\_T \frac{\d\varphi}{\d\phi} = \frac{n\_T K}{2p\_S}.
}
This amounts to letting the dimensions of the Hilbert subspaces form a geometric progression,
\bel{
  \dim \H\_S = \sqrt{ \dim \H \, \dim \H\_T}.
}
A consequence of this pleasant situation is that there is only one upper bound that needs to be found.

To proceed, define
\bel{
  R_n^{(p\_S)} \equiv \sum_{p = -p\_S}^{p\_S - 1} \left|\qprod p n \right|^2.
}
This is a number between $0$ and $1$ that measures the extent to which the $n$'th eigenstate is localized within $-p\_S \leq p < p\_S$. Said another way, this is a measure of smoothness, as defined using the parameter $p\_S$. If $R_n^{(p\_S)} = 1$, then the state $\qvec n$ remains pure upon smoothing. With the choice \eqref{relation pS nG}, $R^{(p\_S)}_n = 1$ simultaneously ensures that $\qvec n$ is tame, i.e.~that it remains pure upon taming.

\begin{figure}
  \centering
  \includegraphics[width=0.5\textwidth]{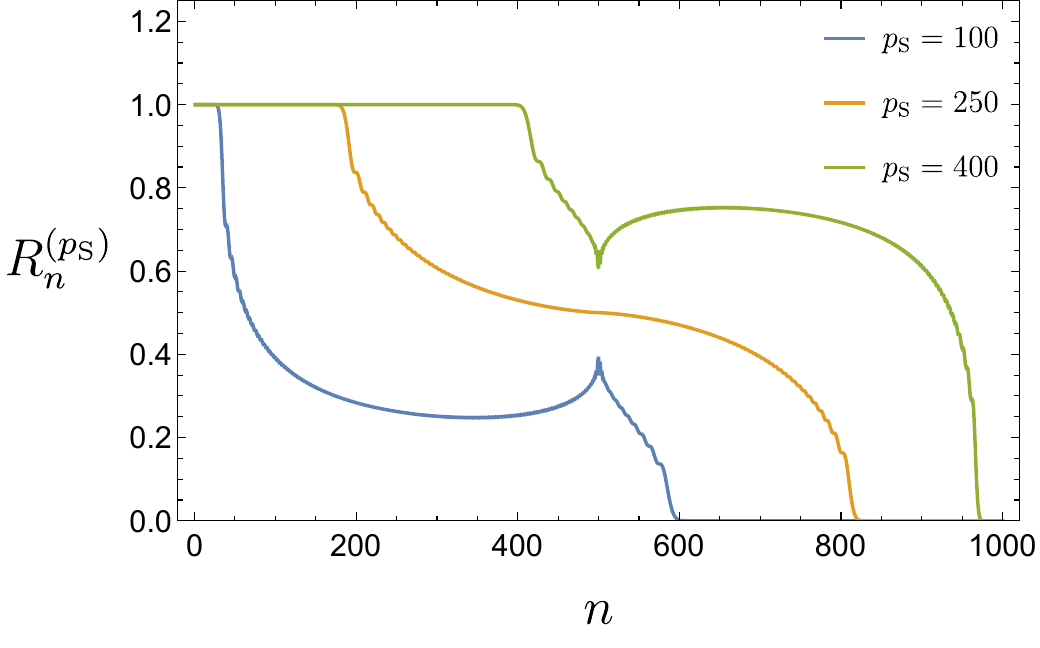}
  \caption{\small The smoothness indicators $R_n^{(p\_S)}$ of all eigenstates $\qvec n$ of $H_g$, for $K = 1000$, $\gamma = 1$, and $n = 1, \ldots, K$, evaluated for multiple values of $p\_S$. The states at $n < K/2$ are, roughly, localized around $\phi = 0$; near $n = K/2$, their wavefunctions acquire a second peak, near $\phi = \pi$, and as $n$ is further increased they become increasingly more centered at this second peak.}\label{fig support}
\end{figure}

The smoothness measures $R_n^{(p\_S)}$ for energy eigenstates are plotted on Fig.\ \ref{fig support}. As the cutoff $p\_S$ is increased past $\sim K/4$, the size of the tame subspace (given by $\dim \H\_T = 2n\_T = (2p\_S)^2/K$) becomes greater than the actual number of tame states in the spectrum (these are states for which $R_n^{(p\_S)} = 1$). After this happens, $\H\_T$ ceases to be an energy eigenspace. Demanding that $\H\_T$ be spanned by energy eigenstates thus places an upper bound on $p\_S$, roughly given by $p\_S \lesssim K/4$. However, since taming already assumes that $p\_S \ll K$, this bound is not very effective. Any sufficiently small $p\_S$, including $p\_S = 100 = K/10$, will automatically have only pure energy eigenstates in $\H\_T$. The upshot, then, is that it is safe to take $p\_S$ to be any number satisfying $\pi/2\sigma \ll p\_S \ll K$: this choice, together with $n\_T = 2p\_S^2/K$, will ensure that the theory
\eqref{def H sho clock} indeed flows to a SHO cQM.

Since $\H\_T$ can be chosen to be an energy eigenspace of $H_g$, the effective Hamiltonian $H'_g$ can be calculated simply by taming $H_g$. 
When acting on smooth states with $|\varphi| \ll \varphi\_T$, $H'_g$ may be expressed as
\bel{
  H_g' \approx -\frac{g^2}{2} \hat\del_\varphi^2 + \frac1{2g^2} \varphi^2.
}
This is almost the familiar SHO Hamiltonian, except $\varphi$ is bounded by $\varphi\_T \ll 1$, while $g^2 = \gamma^2 \d\phi$ vanishes in the $K \gg 1$ limit. These two issues are addressed in one fell swoop. Define
\bel{
  \varphi\_c \equiv \frac{\varphi}{\sqrt{\d\phi}} \quad\trm{and}\quad H_\gamma\^c \equiv -\frac{\gamma^2}2 \hat\del_{\varphi\_c}^2 + \frac1{2\gamma^2} \varphi\_c^2.
}
By choosing parameters so that $\varphi\_T \gg \sqrt{\d\phi}$, $H_\gamma\^c \approx H_g'$ is found to act on smoothly varying states labeled by a continuous parameter $\varphi\_c$ in the region $|\varphi\_c| \ll \varphi\_T/\sqrt{\d\phi}$. Further, $H_\gamma\^c$ has a single $O(1)$ coupling, $\gamma$. It thus describes a cQM of a particle on the line. It is now a simple matter of rescaling by $O(1)$ quantities to express $H_\gamma\^c$ as the usual SHO Hamiltonian.

It may now appear that one could have pursued an alternative --- and more direct --- way of latticizing the SHO Hamiltonian. Consider a lattice theory defined in terms of position and momentum operators \eqref{def P} and \eqref{def Q},
\bel{\label{def H sho PQ}
  H_\omega = \frac12 P^2 + \frac{\omega^2}2 Q^2.
}
Since $P$ and $Q$ tame to the familiar continuum operators, this might seem to be the best pick for a lattice theory that flows to an SHO at low energies. However, this choice is more problematic than the original one, \eqref{def H sho clock}. To see why, first rewrite \eqref{def H sho PQ} in terms of the clock and shift operators, getting
\bel{
  H_\omega = \frac1{2(\d\phi)^2} \left(2 - X^2 - X^{-2}\right) + \frac{\omega^2}{2} \left(2 - Z^2 - Z^{-2} \right).
}
When $K$ is even, this theory is symmetric under the individual maps $X \mapsto -X$ and $Z \mapsto -Z$, generated by the commuting operators $Z^{K/2}$ and $X^{K/2}$. When $K$ is odd, the symmetries are approximate, with corrections of order $1/K$. As a consequence, the system now factors into \emph{four} equivalent SHOs with an effective target regulator $K/4$. Thus all states in the spectrum are (either exactly or up to $1/K$ terms) fourfold degenerate. The degenerate states are related by a shift $\phi \mapsto \phi + \pi$ and by turning on a $(-1)^\phi$ modulation in the target space.

A linear low-energy spectrum is found when $\omega \approx K/{2\pi}$.  Only a quarter of the low-energy wavefunctions are tame and centered around $\phi = 0$. The rest are either centered around $\phi = \pi$, have a $(-1)^\phi$ modulation, or both. This means that the na\"ive taming of the seemingly natural Hamiltonian \eqref{def H sho PQ} will \emph{miss $75\%$ of the low-energy states}. In other words, to get the low-energy physics right in an effective cQM describing the microscopic model \eqref{def H sho PQ}, one \emph{must} work with other, potentially nongaussian sectors. (Alternatively, one must introduce superselection rules that project to individual sectors of the two $\Z_2$ symmetries.) Adding the $\phi = \pi$ sector is the analog of including a soliton in a scalar QFT, and adding the $(-1)^\phi$ modulation is the analog of changing the spin structure of the target space.\footnote{It might seem outlandish to discuss spin structures in a purely scalar theory. However, from a lattice point of view, a choice of a spin structure is simply a choice of how a given subsector of the Hilbert space behaves under a $\Z_2$ symmetry that corresponds to a translation halfway around a (periodic) lattice \cite{Radicevic:2018okd}. Concretely, the theory \eqref{def H sho PQ} has a $\Z_2$ symmetry $X^{K/2}$ that corresponds to a shift by $K/2$ sites, i.e.\ to a translation by $\pi$ along the original target circle. States in different superselection sectors of this symmetry can be represented by wavefunctions on a new, smaller circle, with the two sectors corresponding to the choice of periodic vs.\ antiperiodic boundary conditions. In this sense, equating the $(-1)^\phi$ modulation with changing the spin structure is only the faintest abuse of nomenclature. Indeed, the existence of other low-energy modes can be understood as ``fermion doubling'' without any fermions!}

This analysis holds very broadly. The salient fact in all cases is that a cQM, if it describes the low energies of a lattice theory at all, must break down at high energies, and that this transition is governed by the scale $p\_S$ (and also by $n\_T$ when the cQM target is noncompact).

\section{Smooth path integrals} \label{sec path int}

The story so far has emphasized three ideas: continuum theories are obtained by coarse-graining lattice theories in a specific way; this procedure introduces a hierarchy of scales \eqref{scale hierarchy} in order to define the continuum; and the lattice Hamiltonian imposes bounds on possible choices of coarse-graining scales by demanding that these reductions constitute a flow to a well defined theory. These points can all be formulated in the path integral language. This provides an alternative perspective on smoothing and taming, as well as on the breakdown of the continuum description when the energies become comparable to the various cutoff scales. In addition, this approach provides a rigorous definition of path integrals for continuum theories, and allows certain subtleties to be clearly highlighted.

This Section will focus on deriving and evaluating a path integral expression for the thermal partition function
\bel{\label{def Zf}
  \Zf \equiv \Tr\, \e^{-\beta H}, \quad \beta \in \R^+.
}
The idea behind the derivation is completely standard: $\e^{-\beta H}$ will be expressed as a product of many operators $\e^{-\d\tau H}$ for $\d\tau \ll \beta$, and decompositions of unity will be inserted between them. There are two deviations from the textbook treatment, however:
\begin{enumerate}
  \item Instead of inserting a complete set of states --- or an overcomplete set, as done with coherent state path integrals --- here the goal is to use just a decomposition into smooth (or tame) states. In other words, the novelty here is that an \emph{undercomplete} set of special states is inserted at each point, so only trajectories of a specific jaggedness are retained. This procedure is correct only at sufficiently low temperatures, and with the right relations between various scales involved. This will be the topic of Subsection \ref{subsec def path int}, with some subtler results relegated to Subsection \ref{subsec bounds}.
  \item To actually \emph{calculate} the path integral, further manipulations are necessary. They are not controlled approximations. Their issues will be discussed in Subsection \ref{subsec eval path int}.
\end{enumerate}

\subsection{Constructing the smooth path integral} \label{subsec def path int}

The standard approach to constructing a path integral is to express \eqref{def Zf} as
\bel{\label{Zf as trace}
  \Zf = \Tr \prod_{\tau = \d\tau}^{\beta} \e^{-\d\tau H} = \sum_{\{\greek f_\tau\}} \prod_{\tau = \d\tau}^{\beta} \qmat{\greek f_{\tau + \d\tau}}{\e^{-\d\tau H}}{\greek f_\tau}, \quad \d\tau \equiv \frac \beta {N_0}, \quad \greek f_{\beta + \d\tau} \equiv \greek f_{\d\tau}.
}
for any orthonormal basis $\{\qvec {\greek f}\}$ of $\H$ whose elements are labeled by an ordered set $\{\greek f\}$. An obvious choice here are the clock eigenstates $\{\qvec{\e^{\i\phi}}\}$ labeled by $\greek f = \phi = n \d\phi$ for $1 \leq n \leq K$.

At this point the number $N_0$ is arbitrary, and can even be chosen to be $N_0 = 1$. However, when $\d\tau$ is sufficiently small, the matrix elements $\qmat{\greek f_{\tau + \d\tau}}{\e^{-\d\tau H}}{\greek f_\tau}$ can be approximated with simple functions $\e^{-\d \tau L(\greek f_\tau, \greek f_{\tau + \d\tau})}$. Taking the product over $\tau$ then gives the exponential of the Euclidean action $S[\greek f]$,
\bel{
  \e^{-S[\greek f]} = \e^{-\sum_{\tau = \d\tau }^{\beta} \d\tau L(\greek f_\tau, \greek f_{\tau + \d\tau})}
}
that weights the configurations $\{\greek f_\tau\}$  in just the right way to give the correct $\Zf$. The usefulness of path integrals comes precisely from the fact that by choosing $N_0 \gg 1$, the Lagrangian $L(\greek f_\tau, \greek f_{\tau + \d\tau})$ becomes appealingly easy to handle. This standard story now takes a new turn.

In view of the smoothing philosophy, it is natural to take $\{\qvec{\greek f}\}$ to consist of smooth states $\{\qvec{\e^{\i\varphi}}\}$ and of high-momentum states $\{\qvec p\}_{p\_S \leq p < K - p}$. This is very different from the conventional usage of only clock eigenstates for the $\qvec{\greek f}$'s. The approach chosen here is a simple way to build the target space smoothness of the fields directly into the path integral.

Concretely, the idea is to insert the complete-basis decomposition
\bel{
  \1 = \sum_\varphi \qproj{\e^{\i\varphi}}{\e^{\i\varphi}} + \sum_{p = p\_S}^{K - p\_S - 1} \qproj pp
}
in between the $\e^{-\d\tau H}$'s in \eqref{Zf as trace}. The variable $\varphi$ takes $2p\_S$ different values, for instance $\d\varphi, 2\d\varphi, \ldots, 2\pi$, while $p$ only takes on the ``high'' momentum values (keeping in mind that $\qvec p = \qvec{p + K}$). The key step is to then \emph{drop} all terms involving the high-momentum states. If this approximation is justified, $\Zf$ will end up expressed as a sum over field configurations $\{\varphi_\tau\}$ that encode smeared position eigenstates on the target space. This maneuver must be justified in each microscopic theory separately; for the rest of this Subsection, it will be taken for granted. Note that nothing done so far requires the $\varphi_\tau$'s to be smooth functions of $\tau$.

The next approximation arises when defining the Lagrangian and is actually rather conventional. For simplicity, assume that $H$ has been shifted by a constant so that all energies are positive, and let $\E > 0$ be the largest eigenvalue of $H$.  One can write the operator expression
\bel{
  \e^{-\d\tau H} \approx \1 - H \d\tau
}
only if $\d\tau \ll 1/\E$. If $\d\tau$ is not in this regime, the action will receive corrections from higher powers of $\d\tau$ when describing high-energy states.  At very high temperatures $\beta^{-1} \gg \E$, $\d\tau$ will be in the correct regime for any $N_0$. At lower temperatures, $\beta^{-1} \sim \E^0$, one must have
\bel{
  N_0 \gg \E.
}
If the goal is to reproduce the path integral with only the smooth modes, however, this bound can be relaxed to involve just $\E\_S$, the largest energy scale associated to smooth states.

The third and final set of approximations comes from evaluating the matrix element
\bel{
  \qmat{\e^{\i\varphi_{\tau + \d\tau}}}{ H}{\e^{\i\varphi_\tau}}
}
and obtaining a tractable expression for the Lagrangian $L(\varphi_\tau, \varphi_{\tau + \d\tau})$. This typically involves performing an auxiliary sum over momenta, and this is where approximations need to be invoked with some care.

It is helpful to study these steps in a concrete setup. Consider the free theory \eqref{def H free clock}. In this situation, matrix elements satisfy
\bel{
  \qmat{p}{\e^{- \d\tau H}}{\e^{\i\varphi}} = 0, \quad p\_S \leq p \leq K - p\_S - 1,
}
for any smooth state $\qvec{\e^{\i\varphi}}$. This means that the partition function exactly splits into a sum over low-momentum and high-momentum partition functions. In accordance with the first approximation above, the idea is to focus only on the low-momentum partition function,
\bel{\label{def ZfS}
  \Zf\_S \equiv \sum_{\{\varphi_\tau\}} \prod_{\tau = \d\tau}^{\beta} \qmat{\e^{\i\varphi_{\tau + \d\tau}}}{\e^{-\d\tau H}}{\e^{\i\varphi_\tau}}.
}

The largest energy scale associated with smooth states in this case is $\E\_S \sim p\_S^2$. This means that the second approximation, the expansion in powers of $\d\tau$, gives
\algns{\label{mat element}
  \qmat{\e^{\i\varphi_{\tau + \d\tau}}}{\e^{-\d\tau H}}{\e^{\i\varphi_\tau}}
  &= \frac1{2p\_S} \sum_{p,\, p' = -p\_S}^{p\_S - 1} \e^{\i(p' \varphi_{\tau + \d\tau} - p\varphi_\tau)} \qmat{p'}{\left(\1 - H \d\tau + O(\E\_S^2 \d\tau^2)\right) }{p} \\
  &\hspace{-2em} = \frac1{2p\_S} \sum_{p = -p\_S}^{p\_S - 1} \e^{\i p (\varphi_{\tau + \d\tau} - \varphi_\tau)} \left[1 - \frac{p^2 \d\tau}2  \Big(1 + O \left( p\_S^2\big/ K^2\right)\Big) + O\left(p\_S^4 \d\tau^2 \right) \right] \\
  &\hspace{-2em} \approx \frac1{2p\_S} \sum_{p = -p\_S}^{p\_S - 1} \e^{\i p (\varphi_{\tau + \d\tau} - \varphi_\tau) - \frac12 p^2 \d\tau}.
}
In general, two separate relations need to hold to justify this approximation, $p\_S \ll K$ and $p\_S^2 \d\tau \ll 1$. However, the second requirement can actually be \emph{dropped} in this case because the theory is free; the exponent receives no $p\_S^2 \d\tau$ corrections. In fact, in what follows it will be convenient to assume that $p\_S^2 \d\tau \gtrsim 1$ in order to get familiar Gaussian integrals.

Conventionally, a decomposition of unity in terms of $K$ momentum eigenstates is inserted into the path integral by hand, as a trick that allows the matrix elements to be easily computed. Here, instead, the sum over $2p\_S$ momenta arises from expressing the smooth clock states in terms of shift eigenstates.

The next step is to sum over $p$ while employing the third and final approximation. Define
\bel{
  \alpha^2 \equiv \frac{\d\tau}{(\d\varphi)^2},
}
and let $z \equiv p \sqrt{\d\tau}$ be a new summation variable that changes in steps of $\d z \equiv \sqrt{\d\tau}$. Using this, rewrite the final line of \eqref{mat element} as
\bel{\label{momentum integral}
   \qmat{\e^{\i\varphi_{\tau + \d\tau}}}{\e^{-\d\tau H}}{\e^{\i\varphi_\tau}} = \frac{1}{2\pi\alpha} \sum_{z = -\pi\alpha}^{\pi\alpha - \d z} \d z\, \e^{\i z \frac{\Delta_\tau \varphi}{\sqrt{\d\tau}} - \frac12 z^2}.
}
Note that $|\Delta_\tau \varphi| \equiv |\varphi_{\tau + \d\tau} - \varphi_\tau|$ ranges from $0$ to $\pi$ in steps of $\d\varphi$. If it is close to zero, and if $\alpha \gtrsim 1$, the sum in \eqref{momentum integral} can be approximated by the standard Gaussian integral $\int_{-\infty}^\infty \d z \, \e^{-z^2/2} = \sqrt{2\pi}$. If $|\Delta_\tau \varphi|$ is maximal, e.g.~if $\varphi_{\tau + \d\tau} = \varphi_\tau + \pi$, at each step $\d z$ the phase of the summand will change by precisely $\pi$, while its modulus will remain approximately constant. This means that the sum is well behaved in this extreme limit of $\Delta_\tau \varphi$ (the oscillations will cancel), and so it is reasonable to approximate it by a Gaussian integral for all configurations $\varphi_\tau$. This integral is easily solved at each time step separately, giving
\bel{\label{proto action}
  \Zf\_S \approx \frac1{(2\pi \alpha^2)^{N_0/2}} \sum_{\{\varphi_\tau\}} \e^{- \frac12 \sum_{\tau = \d\tau}^{\beta} \frac{(\Delta_\tau \varphi \, \trm{mod} \, 2\pi)^2}{\d\tau}}.
}

The periodicity of $\varphi_\tau$ is \emph{crucial}. It did not need to be highlighted in \eqref{momentum integral} because $z/\sqrt{\d\tau}$ was an integer, by construction. The emphasis is necessary in \eqref{proto action}. In order to remove it, the range of $\Delta_\tau \varphi$ must be explicitly chosen. There is a simple way to do this: demanding that the symmetry $\varphi_\tau \mapsto -\varphi_\tau$ be na\"ively implemented forces this range to be $-\pi \leq \Delta_\tau \varphi < \pi$. With this implicit choice for what $\Delta_\tau \varphi$ means, the smooth partition function can finally be written in the completely familiar form
\bel{\label{conventional free action}
  \Zf\_S \approx \frac1{(2\pi \alpha^2)^{N_0/2}} \sum_{\{\varphi_\tau\}} \e^{- \frac12 \sum_{\tau = \d\tau}^{\beta} \frac{(\Delta_\tau \varphi)^2}{\d\tau}} \equiv \int [\d\varphi] \, \e^{-\frac12 \int_0^\beta \d\tau \left(\del_\tau \varphi \right)^2}, \quad \del_\tau \varphi \equiv \frac{\Delta_\tau \varphi}{\d\tau}.
}

The conventionality of this result should not distract from the novelties in its derivation. The introduction of the smoothness scale $p\_S$ was key to controlling the approximations that lead to this expression. (At the last step, when converting the sum over $p$ to an integral, it takes a bit more work to give closed expressions for higher-order corrections, but the definition of the Riemann integral makes it clear that the corrections are $O(\d z^2)$ and $O(1/\alpha)$, with the latter ones explicitly involving $p\_S$.) In the following Subsections, it will become clear that $p\_S$ and $n\_T$ are also needed to control the approximations needed when \emph{calculating} the path integral \eqref{conventional free action}.

\newpage

\subsection{Three calculational techniques} \label{subsec eval path int}

The entire song and dance of the previous Subsection is not needed to evaluate the partition function \eqref{def ZfS}. Since the Hamiltonian \eqref{def H free clock} is easy to diagonalize, the smooth partition function can be simply expressed in the energy eigenbasis to give
\bel{\label{part fn boson QM}
  \Zf\_S = \sum_{p = -p\_S}^{p\_S - 1} \e^{-\beta E_p} \stackrel{K \gg p\_S}\approx \sum_{p = -p\_S}^{p\_S - 1} \e^{-\frac12 \beta p^2} \stackrel{1 \gg \beta \gg 1/p\_S^2}\approx \sqrt{\frac{2\pi}\beta}.
}
The final approximation deserves some explanation: if $\beta \gtrsim 1$, the temperature is smaller than the typical level spacing, even at low energies, and so only the ground state contributes to give $\Zf\_S \approx 1$; if $\beta \lesssim 1/p\_S^2$, all states contribute equally, giving $\Zf\_S \approx 2p\_S$; and it is only in between these two extremes that the sum in \eqref{part fn boson QM} can be approximated as a Gaussian integral whose evaluation yields the stated result. This exact result provides a valuable sanity check: the path integral \eqref{conventional free action} must be able to reproduce this quantity. 

As already hinted, trying to evaluate \eqref{conventional free action} comes with its own subtleties. There are at least two different ways to evaluate this path integral, and they both give the correct answer \eqref{part fn boson QM}, though one of them only works with important caveats.

The \emph{direct approach} is to integrate out the variables $\varphi_\tau$ one at a time. A typical summation of this sort would first evaluate
\bel{\label{path int method 1}
  \frac1{\sqrt{2\pi \alpha^2}} \sum_{\varphi_\tau = -\pi}^{\pi - \d\varphi} \e^{-\frac1{2\d\tau} (\varphi_\tau - \varphi_{\tau + \d\tau} \, \trm{mod}\, 2\pi)^2 -\frac1{2\d\tau} (\varphi_\tau - \varphi_{\tau - \d\tau}\, \trm{mod}\, 2\pi)^2}
}
for fixed $\varphi_{\tau \pm \d\tau}$. The goal is to reduce this to a Gaussian integral, just like in the case of the momentum sum \eqref{momentum integral}, and to use the neat identity
\bel{\label{integral identity}
  \int \d z\, \e^{-a (x - z)^2 - b(z - y)^2} = \sqrt{\frac{\pi}{a + b}} \e^{-\frac{ab}{a+b} (x - y)^2}.
}

The sum \eqref{path int method 1} can be replaced by a Gaussian integral when $\alpha \gtrsim 1$. The summand has the form $\e^{-(n^2 + m^2)/2\alpha^2}$, where $n, m \in \Z$ count the number of steps $\d\varphi$ between $\varphi_\tau$ and $\varphi_{\tau \pm \d\tau}$, and $\alpha \gtrsim 1$ ensures that large values of $n$ and $m$ are suppressed. The sum is thus dominated by the term(s) in which $\varphi_\tau$ is close to $\varphi_{\tau + \d\tau}$ and $\varphi_{\tau - \d\tau}$. This means that the $2\pi$ periodicity in \eqref{path int method 1} can be ignored, as it is only relevant when differences between $\varphi_\tau$ and $\varphi_{\tau + \d\tau}$ are maximal --- and such configurations are strongly suppressed in the sum. Thus the sum \eqref{path int method 1} is equal to
\bel{
  \frac1{\sqrt{2}} \e^{-\frac1{4\d\tau} \left(\varphi_{\tau + \d\tau} - \varphi_{\tau - \d\tau} \right)^2}.
}

Repeating the process for all the $\varphi_\tau$'s except the last one gives
\bel{\label{path int result 1}
  \Zf\_S \approx \frac1{\sqrt{2\pi N_0 \alpha^2}} \sum_{\varphi_\beta = \d\varphi}^{2\pi} \e^{-\frac{(\Delta\varphi)^2}{2N_0 \alpha^2}} = \frac1{\sqrt{2\pi N_0 \alpha^2}} 2p\_S =\frac{2p\_S\d\varphi}{\sqrt{2\pi \beta}} = \sqrt{\frac{2\pi}\beta}.
}
In this calculation, $\Delta \varphi \equiv \sum_\tau \Delta_\tau \varphi$ is the phase accumulated while winding around the thermal circle; this has to be an integer multiple of $2\pi$, and by the convention that $|\Delta\varphi| \leq \pi$ in all path integrals, this number can be set to zero. The rest of the calculation is then straightforward, and the correct answer \eqref{part fn boson QM} is obtained.

The direct approach also clarifies why the path integral \eqref{conventional free action} fails if $\beta \lesssim 1/p\_S^2$. This relation can be equivalently written as
\bel{
  N_0 \d\tau \lesssim (\d \varphi)^2.
}
It thus implies
\bel{
  \alpha^2 \lesssim \frac1{N_0} \ll 1,
}
and so it is incompatible with the assumption $\alpha \gtrsim 1$ that was crucial in the above derivation.

The \emph{frequency space approach} to calculating \eqref{conventional free action} is more involved than the direct one, but it has the advantage of generalizing more readily to other situations. The basic idea is familiar: assume for simplicity that $N_0$ is even, and perform the Fourier transform
\bel{\label{def varphi n}
  \varphi_\tau \equiv \frac1{\sqrt{N_0}} \sum_{n = - \frac12 N_0}^{\frac12 N_0 - 1} \varphi_n \, \e^{\i \omega_n \tau},\quad  \omega_n \equiv \frac{2\pi}{\beta} n.
}
The quantities $\omega_n$ are the usual Matsubara frequencies. Then, the na\"ive story goes, the path integral \eqref{conventional free action} can be approximated by a product of ordinary Gaussian integrals, each over one variable $\varphi_n$.

Unfortunately, this is just a convenient fantasy. The inconvenient truth is that the periodicity of $\varphi_\tau$ causes the modes $\varphi_n$ to take values in $n$-dependent sets: the sums over them are not Gaussian integrals. This reflects the fact that no eigenstates of the free clock Hamiltonian are tame states.

There exists a well known refinement of the above na\"ive story. In order to be in a regime in which the momentum modes can be viewed as variables of independent Gaussian integrals, the variables $\varphi_\tau$ need to be noncompact, i.e.\ it must be possible to approximate the full path integral with one in which the $\varphi_\tau$'s can be somehow restricted to range over only a small subset of their possible values. The hope is that this would give a doable calculation while still approximately calculating $\Zf\_S$.

It is not possible to restrict \emph{all} the $\varphi_\tau$'s to be small and get a good approximation to $\Zf\_S$. However, the path integral variables can be restricted in a more oblique way. Divide the set of $2p\_S$ values that each $\varphi_\tau$ can take into $p\_S/n\_T$ overlapping subsets, each with $2n\_T$ sequential values of $\varphi_\tau$. For each $\tau$, restrict the variable $\varphi_\tau$ to just one of these subsets. In other words, write
\bel{\label{def delta phi}
  \varphi_\tau \equiv \varphi_\tau\^{cl} + \delta\varphi_\tau,
}
where $\delta\varphi_\tau$ are the variables of integration that satisfy
\bel{
  -\varphi\_T \leq \delta\varphi_\tau < \varphi\_T, \quad \d\varphi \ll \varphi\_T \ll 1,
}
and where $\varphi_\tau\^{cl}$ are ``classical'' or ``background fields'' --- mere labels for the particular interval to which $\varphi_\tau$ is restricted.  It is not sufficient to restrict all the $\varphi_\tau$'s to the same interval --- instead of working just with $\varphi_\tau\^{cl} = 0$, other backgrounds must be included.

To justify restrictions of the form \eqref{def delta phi} and to find what backgrounds $\varphi_\tau\^{cl}$ need to be included, consider the action $S[\varphi]$. When a field varies by a lot between neighboring points, i.e.\ when $|\Delta_\tau\varphi| \sim 1$, the action gets a contribution of size $1/\d\tau$. When all fluctuations are small, say $|\Delta_\tau \varphi| < \varphi\_T \equiv n\_T \d\varphi$ for all $\tau$, the action is
\bel{
  S \lesssim N_0 \frac{n\_T^2 \d\varphi^2}{\d\tau} = N_0 \frac{n\_T^2}{\alpha^2}.
}
This is a rough indication that there exists a parameter regime in which even a single large fluctuation is sharply suppressed in the path integral, and this regime is
\bel{\label{relation alpha beta}
  \frac1{\d\tau} \gg \frac{N_0 n\_T^2}{\alpha^2} \Longleftrightarrow \alpha^2 \gg n\_T^2 \beta.
}
This  can also be written as
\bel{\label{alpha bound}
  \frac{p\_S^2}{n\_T^2} \gg N_0.
}
Note that this is consistent with both $1 \gg \beta \gg 1/p\_S^2$ and with $\alpha^2 \gtrsim 1$.

In this parameter regime, if $\varphi_\tau$ does make a full $2\pi$ winding as it varies along the thermal circle, it must do so slowly, over many steps $\d\tau$. This fact underlies the use of homotopy theory to classify spacetime solitons.

The saddle-point method is a further refinement of this analysis. Concretely, one may focus only on configurations that represent small oscillations around the local minima of $S[\varphi]$. These minima --- the solutions to the equations of motion $\delta S/\delta \varphi = 0$ --- are precisely the background (``classical'') fields $\varphi_\tau\^{cl}$ that one must include in order to use the Fourier transform \eqref{def varphi n} to evaluate \eqref{conventional free action} in a consistent approximation. While this may sound like a tale as old as time, the bounds on its validity presented here, such as \eqref{relation alpha beta}, are new.

\newpage

The equation of motion of the action \eqref{conventional free action} is $\Delta_\tau^2 \varphi = 0$, and its solutions are of the form
\bel{\label{def winding modes}
  \varphi_\tau\^{cl} = \varphi\^{cl}_0 + \frac{2\pi w}{\beta} \tau,
}
where $w$ is an integer that counts the number of windings of $\varphi_\tau\^{cl}$ around the thermal circle. The maximal value of $w$ depends on the various parameters introduced so far: for instance, if $|\Delta_\tau \varphi| < 2\varphi\_T$ is required, then  the only winding sectors that can be included have
\bel{\label{winding limits}
  |w| \lesssim N_0 \varphi\_T \ll \frac{p\_S}{n\_T}.
}

\emph{Important aside}: In the canonical formalism, the taming cutoff $\varphi\_T$ had no r$\hat{\trm o}$le to play in the free clock model. This cutoff was also not needed to \emph{define} the path integral \eqref{conventional free action}. Right now, its only use is to facilitate a particular approximation of this path integral that exploits powerful frequency space methods. If the underlying theory were tamed, the very definition of the corresponding path integral over tame states would have depended on $\varphi\_T$, and the integral would have featured up to $p\_S/n\_T$ $\tau$-independent sectors without overlap, each encompassing $2n\_T$ sequential positions $\varphi$. For example, the path integral that computes a partition function $\Zf\_T$ of tame states in the theory \eqref{def H sho PQ} would feature two sectors, one with states localized around $\varphi\^{cl} = 0$ and another with states localized around $\varphi\^{cl} = \pi$, and not a single field configuration entering this integral would interpolate between these two sectors the way configurations \eqref{def winding modes} do.

There are thus two logically distinct ways Gaussian integrals arise when doing path integrals: they may appear from restricting to configurations near a saddle point, or they may appear because the underlying theory really has tame (i.e.\ localized) eigenstates. Theories described by the former path integrals are often called \emph{compact}, while the latter path integrals describe \emph{noncompact} theories; the distinguishing feature of compact theories is that their path integrals include configurations that interpolate between different tame sectors. (Non)compactness is sometimes taken to be a rather fundamental property of a theory, but the examples given here illustrate that both compact and noncompact path integrals may arise from lattice theories with the exact same microscopic Hilbert space.

So much for this digression; now it is time to calculate the path integral \eqref{conventional free action}. The upshot of the previous few pages is that the partition function can be approximated by
\bel{
  \Zf\_S \approx \frac1{(2\pi\alpha^2)^{N_0/2}} \sum_{w, \, \varphi\^{cl}_0} \e^{-\frac{(2\pi)^2}{2\beta} w^2} \sum_{\{\delta \varphi_\tau\}} \e^{ - \frac12 \sum_{\tau = \d\tau}^\beta \d\tau \, (\del_\tau \delta\varphi)^2}.
}
The sum over $w$ runs over all integers bounded by \eqref{winding limits}; the $\varphi_0\^{cl}$'s will be discussed later. The sum over fluctuations $\delta\varphi_\tau$, at each point absolutely bounded by $\varphi\_T$, is the main issue.

It is convenient to alter the normalization of the Fourier transform \eqref{def varphi n} when applying it to $\delta\varphi_\tau$, so define
\bel{
  \delta \varphi_\tau \equiv \sum_{n = - \frac12 N_0}^{\frac12 N_0 - 1} \delta\varphi_n \, \e^{\i \omega_n \tau}.
}
In frequency space, the action becomes
\bel{\label{def S}
  S[\delta\varphi] \equiv \frac12 \sum_{\tau = \d\tau}^\beta \d\tau \, (\del_\tau \delta\varphi)^2 = \frac\beta2 \sum_{n = -\frac12 N_0}^{\frac12 N_0 - 1} |\delta\varphi_n^2| \, \frac{4\sin^2\frac{\omega_n \d\tau}2}{(\d\tau)^2}.
}
As promised, the momentum modes can be viewed as small: from $|\delta\varphi_\tau| \leq \varphi\_T$, it follows that
\bel{
  |\delta\varphi_n| \leq \frac1\beta \int_0^\beta \d\tau |\delta\varphi_\tau| \leq \varphi\_T.
}
While tighter bounds exist, this is enough to justify replacing the sum over $\delta\varphi_n$ with a Gaussian integral. The one exception is at $n = 0$; the zero mode simply drops out of the action. (The same happened with $\varphi_0\^{cl}$; indeed, the summations over these two variables can be combined to get a simple factor of $2p\_S$ in front of the integrals.) The sum with the zero mode omitted (denoted by a prime) can then be written as
\bel{
  \sideset{}{'}\sum_{\{\delta \varphi_\tau\}} \e^{ - S[\delta\varphi]} \approx \N \int \d(\delta\varphi_{-N_0/2}) \, \e^{-\beta \eps_{-N_0/2} \delta\varphi_{-N_0/2}^2 } \, \prod_{n = 1}^{\frac12 N_0 - 1} \int \d^2(\delta\varphi_n) \, \e^{-\beta \eps_n |\delta\varphi_n^2| },
}
where $\N \equiv (\d\varphi)^{-N_0 + 1}$, and $\eps_n \equiv \frac4{(\d\tau)^2}\sin^2\frac{\omega_n \d\tau}2$. Doing the Gaussian integrals, and noting that $w = 0$ dominates the sum over windings at $\beta \ll 1$, finally gives
\algns{\label{boson mats prod}
  \Zf\_S
  &\approx \frac{2p\_S \, \N}{(2\pi\alpha^2)^{N_0/2}} \sum_{w} \e^{-\frac{(2\pi)^2}{2\beta} w^2} \sideset{}{'}\prod_n \sqrt{\frac{\pi}{\beta\eps_n}} \approx \frac{\sqrt{4\pi \beta}}{(2N_0)^{N_0/2}} \sideset{}{'}\prod_n \frac{\d\tau}{2|\sin\frac{\omega_n \d\tau}2|}.
}

A standard sequence of uncontrolled approximations can now be used in order to proceed. The first step is to simply \emph{drop} the high frequencies from the product and approximate the remaining terms by $|\omega^{-1}_n|$. The second step is to \emph{extend} the range of $n$ to all the integers, and then to compute $\prod'_{n \in \Z} |\omega_n^{-1}|$ using e.g.\ $\zeta$-function methods, getting the finite (``universal'') answer $1/\beta$ (see e.g.\ \cite{Polchinski:1998rq}). The third step, done in lockstep with the second, is to again \emph{drop} any prefactors that explicitly depend on $N_0$. This ultimately yields
\bel{\label{def tilde Zf}
  \trm{``}\Zf\_S\trm{''} = \sqrt{\frac{2\pi}\beta}.
}

The answer \eqref{def tilde Zf} is in quotation marks because this quantity is fundamentally \emph{different} from the smooth partition function \eqref{conventional free action}. Neither of the three steps in the last paragraph leaves $\Zf\_S$ approximately invariant. The idea is to use Wilsonian universality and view $\Zf\_S$ as a microscopically defined quantity that within itself harbors a universal quantity --- a quantity that does not depend on factors like $N_0$ or $p\_S$ that must be ``taken to infinity'' in order to reach the cQM description. Therefore, universality suggests, any modification of the path integral that only affects large frequencies (and terms that explicitly depend on cutoffs like $N_0$) is acceptable, as it can be expected not to affect the universal, low-frequency behavior.

This expectation is not quite borne out, and the multi-step procedure above must be treated with care. In particular, in order to get the answer \eqref{def tilde Zf}, the prefactor $1/(2N_0)^{N_0/2}$ in \eqref{boson mats prod} had to be  replaced by $\frac1{\sqrt 2}$ in the third step. This replacement can be thought of adding an $N_0$-dependent \emph{counterterm} to the action, in this case $-\frac{N_0}2 \log(2N_0) + \frac12 \log 2$. There is a freedom to change the finite part $\frac12 \log 2$ of the counterterm and thereby change the final (universal) result. Indeed, it is part of renormalization lore that counterterms must be fixed by additional input, e.g.\ by asking that a known answer be reproduced in some special case. The term $\frac12\log 2$ was chosen specifically to reproduce the result \eqref{part fn boson QM}.

One final thought for this Subsection. It is actually the \emph{first} step of the above procedure that truly warrants highlighting here. Restricting to low Matsubara frequencies replaces the action \eqref{def S} by
\bel{
  \~S[\delta\varphi] = \frac{\beta}2 \sum_{n = -n\_S}^{n\_S - 1} \omega_n^2\, |\delta \varphi_n^2|, \quad 1 \ll n\_S \ll N_0.
}
This is the temporal analogue of smoothing (as reflected by the subscript of the newly introduced cutoff $n\_S$). This procedure has no counterpart in the canonical formalism. The action $\~S$ is not a bona fide action of a unitary quantum theory. However, it is precisely this kind of action that describes the commonly used continuum path integrals.

Removing the high-frequency modes defines a map
\bel{\label{temporal smoothing}
  \delta\varphi_\tau \mapsto \delta\varphi(\tau) \equiv \sum_{n = -n\_S}^{n\_S - 1} \delta\varphi_n \, \e^{\i \omega_n \tau},
}
where $\delta\varphi(\tau)$ are path integral variables that vary smoothly along the temporal direction,
\bel{\label{temporal smoothness}
  \delta\varphi(\tau + \d\tau) = \delta\varphi(\tau) + \d\tau\, \hat\del_\tau \delta\varphi(\tau) + O(n\_S^2/N_0^2).
}
Thus \eqref{temporal smoothing} is a way to construct fields that vary smoothly across a discrete spacetime. The price to pay is the need to assume that universality holds (in the sense explained above) and to then determine the right counterterms that must be included into the action.

The temporally smooth action $\~S[\delta\varphi]$ can be recorded in ``position'' space as
\bel{\label{def S tilde}
  \~S[\delta\varphi] = \frac12 \sum_{\tau = \d\tau}^{\beta} \d\tau \big(\del_\tau \delta \varphi(\tau) \big)^2 \equiv \frac12 \int_0^\beta \d\tau \big(\del_\tau \delta \varphi(\tau) \big)^2.
}
The difference from the original action \eqref{def S} is that here the variables are constrained to be smooth functions of the Euclidean time. This is the action that is often taken as the starting point when defining a free scalar cQFT.

To illustrate the difference between \eqref{def S} and \eqref{def S tilde}, consider the behavior of these actions under dilatations. First, define the dilatation of the microscopic field $\delta\varphi_\tau$ as the map
\bel{
  \delta\varphi_\tau \mapsto \lambda^\Delta \delta\varphi_{\lambda \tau}.
}
Even the definition of this transformation is troublesome. The scale parameter $\lambda$ must be an integer, and therefore it is impossible to talk about infinitesimal dilatations. Moreover, the transformation is typically not a bijection on the time circle: when $\lambda = 2$, say, both $\varphi_{\beta/2}$ and $\varphi_\beta$ are mapped to $\varphi_\beta$. There is essentially nothing to be learned from dilatations in the original action $S[\delta\varphi]$.

In constrast, the map
\bel{
  \delta\varphi(\tau) \mapsto \lambda^\Delta \delta\varphi(\lambda \tau)
}
can be meaningfully defined for all $\lambda \lesssim 1$ by using the smoothness property \eqref{temporal smoothness}. Furthermore, the fact that the sum \eqref{def S tilde} runs over all $\tau$'s gives rise to a very useful redundancy. Since $\delta\varphi(\tau + \d\tau) = \delta\varphi(\tau) + O(n\_S/N_0)$, at first order these two time points (and all the others a short distance from $\tau$) contribute equally to $\~S[\delta\varphi]$. This in turn justifies changing variables in the action in the usual way, and then it is possible to see that at $\Delta = -1/2$ the action remains unchanged by the dilatation, except for the rescaling $\beta \mapsto \lambda\beta$. Much more about these manipulations will be said in the next paper of this series.

\subsection{When is summing over smooth trajectories justified?} \label{subsec bounds}

Having evaluated the partition function of smooth states $\Zf\_S$, it is now possible to justify the one approximation from Subsection \ref{subsec def path int} that has not been justified so far: using $\Zf\_S$ in lieu of to the full partition function $\Zf$ in a suitable range of temperatures. Let
\bel{
  \Zf\_{nS} \equiv \Zf - \Zf\_S
}
be the contribution from all field configurations that involve a high-momentum state $\qvec p$, $p\_S \leq p < K - p\_S$, at at least one moment in Euclidean time.

In the free clock model, as discussed earlier, $\Zf\_{nS}$ contains only contributions from field configurations that have high momenta at all times $\tau$. Directly evaluating $\Zf\_{nS}$ is forbiddingly difficult. A simple bound can be established by using the fact that energies of all nonsmooth states are $E_p \geq \frac12 p\_S^2$. This means that their contribution to the total partition function obeys
\bel{
  \Zf\_{nS} \leq (K - 2p\_S) \e^{-\frac12 \beta p\_S^2}
}
at all temperatures. Demanding that this be much smaller than $\Zf\_S \approx \sqrt{2\pi/\beta}$ implies
\bel{\label{beta bound}
  \frac{2\pi} \beta \gg (K - 2p\_S)^2 \e^{-\beta p\_S^2} \implies \beta \gg \frac{2\log K}{p\_S^2}.
}
Recall that the calculation of the smooth partition function \eqref{part fn boson QM} assumed only that $\beta \gg 1/p\_S^2$. The bound \eqref{beta bound} is more restrictive, and this indicates that the smooth partition function $\Zf\_S$ cannot approximate $\Zf$ at all temperatures at which it can be reliably calculated. Note that this bound is very weak: tighter ones must certainly exist, and finding them remains an interesting open question.

Another way to frame this observation is as follows. A sufficiently deep numerical exploration of the lattice theory \eqref{def H free clock} would find a crossover from $\Zf \sim 1/\sqrt \beta$ to $\Zf = K = \dim \H$ as the temperature increased. In particular, the fact that $\Zf \not\approx \Zf\_S$ would have to become obvious at $\beta \sim 1/p\_S^2$, when $\Zf\_S$ must plateau at value $2p\_S$ while $\Zf$ continues to grow all the way to $K$. The bound \eqref{beta bound} says that more is true: even before the temperature was increased to the point $\beta \sim 1/p\_S^2$, the full answer $\Zf$ would start showing signs of a crossover. The behavior at $\beta \sim \log K/p\_S^2$ would thus become contaminated by nonsmooth states. This is reminiscent of the roughening transition found in Monte Carlo studies of lattice gauge theories, where the transition between low- and high-temperature (confining and deconfining) regimes is presaged by a crossover into a regime in which jagged flux lines start dominating the partition function \cite{Itzykson:1980fz}.

The fact that this ``roughening transition'' depends on two separate scales, $K$ and $p\_S$, has an important consequence. If $\log K \gtrsim p\_S^2$, the bound \eqref{beta bound} implies $\beta \gg 1$, which is at odds with the upper bound $\beta \ll 1$ used to get \eqref{part fn boson QM}. This means that $K$ cannot be arbitrarily larger than $p\_S$, or else the nonsmooth states would always dominate in $\Zf$. Thus, demanding that a cQM regime exist at \emph{some} temperatures implies the (again, rather weak) upper bound
\bel{\label{K bound}
  K \ll \e^{p\_S^2/2}.
}
It is fascinating that finite-temperature considerations place a general bound on the possible relation between the two cutoffs needed to define a continuum theory.

\newpage

\section{Fermions} \label{sec ferm}

There are no fermions in QM; the notion of particle statistics does not exist in $(0 + 1)$D. In this paper, a ``fermion'' will be any two-level system, i.e.~a theory with a two-dimensional Hilbert space. This is the minimal Hilbert space in which nontrivial dynamics can happen, and as such it is the polar opposite of the $K \gg 1$ theories studied so far. In particular, there is no way to define fermion smoothing in the canonical formalism. However, two-level systems are exceptional because they admit a natural path integral description (albeit in terms of anticommuting variables) that is in many ways analogous to the one seen in clock models. The notion of temporal smoothing in the path integral, developed in Subsection \ref{subsec eval path int}, is one of the concepts that apply to fermionic path integrals too. This makes fermions an attractive playground in which subtle notions of path integral smoothing and universality can be explored and compared to easily obtained exact answers. The ideas developed here will find further applications in the QFT analyses of other papers in this series.

\subsection{Smooth Berezin integrals} \label{subsec Berezin}

Path integrals for fermions, often called Berezin integrals, are covered in every QFT textbook. This Subsection will thus be rather telegraphic, and in order to liven things up it will be presented from a somewhat unusual point of view. This presentation was inspired by \cite{Creutz:1999zy}.

Start with a two-dimensional Hilbert space $\H$. Its algebra of operators, $\C^{2\times 2}$, is spanned by the identity and the three Pauli matrices, denoted $\sigma^x$, $\sigma^y$, and $\sigma^z$. Fermion annihilation and creation operators are defined, respectively, as
\bel{
  \psi \equiv \frac12(\sigma^x + \i \sigma^y), \quad \psi\+ \equiv \frac12(\sigma^x - \i \sigma^y).
}

A few more definitions are needed now. A \emph{(unital) Grassmann algebra} will here refer to an algebra over the complex numbers generated by $N + 1$ operators $\{\eta_i\}_{i = 1}^N \cup \{\1\}$ satisfying
\bel{
  \{\eta_i, \eta_j\} = 0, \quad [\eta_i, \1] = 0.
}
for all $i$ and $j$. It is convenient to work with $N = 2M$ anticommuting generators (``Grassmann numbers'') $\eta_i$ and $\bar \eta_i$ for $1 \leq i \leq M$. These generators can be represented as annihilation operators in an auxiliary system of $2M$ fermions whose Hilbert space, $\bigotimes_{i = 1}^M (\H_i \otimes \bar\H_i)$, is isomorphic to $\H^{\otimes 2M}$. (Graded products are implied throughout.) Importantly, the Grassmann algebra is not $*$: there is no ``Hermitian conjugation'' relating its generators, so $\eta_i$ and $\bar \eta_j$ have a vanishing anticommutator even at $i = j$. In other words, no operator in this Grassmann algebra acts as a creation operator in the auxiliary system of $2M$ fermions.

There is a standard definition of an integral over the variable $\eta_i$. Any element of the Grassmann algebra can be written as $\mathcal X \equiv \mathcal X^{(i)}\_b + \eta_i \mathcal X^{(i)}\_s$, where $\mathcal X^{(i)}\_{b/s}$  do not involve $\eta_i$. 
Then the Berezin integral is
\bel{\label{Berezin def 1}
  \int \d\eta_i \left(\mathcal X^{(i)}\_b + \eta_i \mathcal X^{(i)}\_s \right) \equiv \mathcal X\_s^{(i)}.
}
Since Grassmann numbers are operators in the auxiliary system of fermions, the integral of $\mathcal X$ over $\eta_i$ can be defined via a trace over the $i$'th auxiliary fermion,
\bel{\label{Berezin def 2}
  \int \d\eta_i\, \mathcal X \equiv \tr_{\H_i} \left(\eta_i\+ \mathcal X\right).
}
(Uppercase ``Tr'' is reserved for traces in the original two-state system.) An integral over $\bar\eta_i$ can likewise be expressed as a multiplication by $\bar\eta_i\+$ followed by a trace over $\bar \H_i$. Note that $\eta_i\+$ and $\bar\eta_i\+$ do \emph{not} belong to the starting Grassmann algebra.

Now consider the original single fermion theory, where any operator can be written as
\bel{
  \O = a\1 + b \psi + b' \psi\+ + c \psi\+ \psi.
}
The trace of this operator, easily calculated to be $\Tr\O = 2a + c$, can also be expressed as
\bel{\label{Berezin trick}
  \Tr \O = \int \d\eta \d\bar\eta\, \e^{2\bar\eta \eta} \mathcal X, \quad \trm{where} \quad \mathcal X \equiv \O\big|_{\psi \mapsto \eta,\  \psi\+ \mapsto \bar \eta} = a\1 + b \eta + b' \bar \eta + c \bar\eta \eta.
}
This follows from a short brute force calculation, using $\e^{2\bar\eta \eta} = \1 + 2\bar \eta \eta$. 

More generally, the product of multiple operators can be expressed as
\bel{\label{Berezin multi trick}
  \Tr\left(\O_1 \cdots \O_{N_0}\right) = \int \d\eta_1 \d\bar\eta_1 \cdots \d\bar\eta_{N_0} \, \e^{\bar \eta_1 (\eta_1 + \eta_{N_0})} \e^{\bar \eta_2 (\eta_2 - \eta_1)} \cdots \e^{\bar \eta_{N_0} (\eta_{N_0} - \eta_{N_0 - 1})} \mathcal X_1 \cdots \mathcal X_{N_0}.
}
This trick is easily proven by induction. Applying it to a fermion partition function gives
\bel{\label{def Z one fermion}
  \Zf = \Tr \, \e^{-\beta H} = \Tr \prod_{n = 1}^{N_0} \e^{-\d\tau H} \approx \int \d\eta_{\d\tau} \d\bar\eta_{\d\tau} \cdots \d\bar\eta_\beta \, \e^{\sum_{\tau = \d\tau}^{\beta} \d\tau \left[ \bar \eta_\tau (\del \eta)_{\tau - \d\tau} - H(\eta_\tau, \bar\eta_\tau)\right]}
}
with antiperiodic boundary conditions $\eta_0 \equiv - \eta_\beta$. The approximation comes from assuming $\d\tau$ to be infinitesimal; at a general (possibly small) $N_0$,  $\e^{-\d\tau c \psi\+\psi}$ should be replaced with $\1 + (\e^{-c \d\tau} - 1)\bar\eta \eta$. The discrete derivative is defined as in \eqref{conventional free action}, $(\del \eta)_\tau \equiv \frac1{\d\tau}(\eta_{\tau + \d\tau} - \eta_\tau)$.

There is no regime in which the Grassmann variables $\eta_{\tau - \d\tau}$ and $\eta_\tau$ are ``close'' to each other. In fact, the present approach makes it clear that they should be thought of as \emph{fixed} matrices with eigenvalues $\pm 1$. Nevertheless, they can be transformed to frequency space, and then the three-step procedure of Subsection \ref{subsec eval path int} can be applied to smooth them out.

As a first step towards smoothing, define the Fourier transforms
\bel{\label{def eta n}
  \eta_\tau \equiv \frac1{\sqrt{N_0}} \sum_{n = -\frac12{N_0}}^{\frac12{N_0} - 1} \eta_n\, \e^{\i \omega_n \tau}, \quad \bar \eta_\tau \equiv \frac1{\sqrt{N_0}} \sum_{n = -\frac12{N_0}}^{\frac12{N_0} - 1} \bar \eta_n\, \e^{- \i \omega_n \tau},
}
with Matsubara frequencies
\bel{
  \omega_n \equiv \frac{2\pi}{\beta} \left(n + \frac12\right)
}
whose half-integer offsets reflect the antiperiodic boundary conditions. The Euclidean action
\bel{\label{def S ferm}
  S[\eta, \bar\eta] = \sum_{\tau = \d\tau}^{\beta} \d\tau \big[ -\bar \eta_\tau (\del \eta)_{\tau - \d\tau} + H(\eta_\tau, \bar\eta_\tau) \big]
}
has the kinetic term that becomes, in frequency space,
\algns{\label{Berezin kinetic}
  -\sum_{\tau = \d\tau}^{\beta} \d\tau\, \bar\eta_\tau (\del\eta)_{\tau - \d\tau}
  = - \sum_{n = -\frac12N_0}^{\frac12N_0- 1} \!\! \bar\eta_n \eta_n \left(1 - \e^{-\i \omega_n \d\tau} \right).
}
The most general potential term is $H(\eta_\tau, \bar\eta_\tau) = h_0 \1 + h_1 \bar\eta_\tau \eta_\tau$, which translates to
\bel{\label{Berezin potl}
  \sum_{\tau = \d\tau}^{\beta} \d\tau \, H(\eta_\tau, \bar\eta_\tau) = \beta h_0 + h_1 \d\tau \sum_{n = -\frac12{N_0}}^{\frac12N_0- 1} \bar\eta_n \eta_n.
}

Recall that the Berezin integral can be interpreted as a trace over $2M$ auxiliary fermions, with each $\eta_\tau$ and $\bar\eta_\tau$ acting as an annihilation operator. Remarkably, the Berezin integral can also be viewed as a trace over $2M$ auxiliary fermions \emph{in frequency space}. The reason is that $\eta_n$ and $\bar\eta_n$ also anticommute and square to zero, and can thus be interpreted as bona fide fermion annihilation operators on their own. 
Integrating them out, using the actions \eqref{Berezin kinetic} and \eqref{Berezin potl}, gives
\bel{\label{Matsubara product}
  \Zf \approx \e^{-\beta h_0} \prod_{n = -\frac12{N_0}}^{\frac12{N_0} - 1} \left(1 - \e^{-\i \omega_n \d\tau} - h_1\d\tau \right).
}

The partition function can also be directly evaluated by taking the trace over $\H$. The result is
\bel{
  \Zf = \e^{-\beta h_0} \left(1 + \e^{-\beta h_1} \right).
}
It is easy to numerically check that the product in \eqref{Matsubara product} indeed equals $1 + \e^{-\beta h_1}$ at $h_1 \d\tau \ll 1$. Delicate phase cancellations in the high-frequency terms are crucial to get this result.

Now recall that the smoothing procedure of Subsection \ref{subsec eval path int} hinged on the restriction \eqref{temporal smoothing} to low Matsubara frequencies. Consider the analogous map of Grassmann variables
\bel{\label{temporal smoothing ferm}
  \eta_\tau \mapsto \eta(\tau) \equiv \frac1{\sqrt{N_0}} \sum_{n = -n\_S}^{n\_S - 1} \eta_n \, \e^{\i\omega_n \tau}, \quad   \bar\eta_\tau \mapsto \bar\eta(\tau) \equiv \frac1{\sqrt{N_0}} \sum_{n = -n\_S}^{n\_S - 1} \bar\eta_n \, \e^{-\i\omega_n \tau}.
}
This normalization ensures that the frequency-space modes obey the canonical anticommutation relations  $\{\eta_n, \eta_m\+\} = \{\bar\eta_n, \bar\eta_m\+\} = \delta_{nm}$. The smeared variables $\eta(\tau)$ satisfy the now-familiar smoothness relation of the form
\bel{
  \eta(\tau + \d\tau) = \eta(\tau) + \d\tau \, \hat\del_\tau \eta(\tau) + O(n\_S^2/N_0^2).
}
Dropping the high frequencies also changes the action, with the new one being
\bel{\label{def S tilde ferm}
  \~S = \beta h_0 + \sum_{n = -n\_S}^{n\_S - 1} \d\tau \left(-\i\omega_n  + h_1\right) \bar\eta_n \eta_n.
}
This action in turn defines a new partition function
\bel{\label{def Z tilde ferm}
  \~\Zf \equiv \int [\d\eta\d\bar\eta] \, \e^{-\~S[\eta, \bar\eta]} = \e^{-\beta h_0} \prod_{n = -n\_S}^{n\_S - 1} \left(\i \omega_n - h_1 \right) \d\tau,
}
where $[\d\eta\d\bar\eta] \equiv \prod_{n = -n\_S}^{n\_S - 1} \d\eta_n \d\bar\eta_n$.\footnote{There is some freedom here to change definitions. For instance, including higher frequencies in the measure --- but not in the action --- would simply change the multiplicative prefactor of the partition function, and this change could be compensated by changing the counterterm appropriately.} As in the scalar case, in no way does the restriction to low frequencies approximate the exact result. As $N_0$ is taken to infinity while $\beta$ and the $h$'s remain $O(1)$, $\~\Zf$ vanishes, which means that $\~\Zf$ is not a physical partition function.

With suitable processing, it is possible to extract the correct result from $\~\Zf$. The usual $\zeta$-function or Pauli-Villars methods (see e.g.\ \cite{Headrick:2008ke}) reveal that the universal part of the product \eqref{def Z tilde ferm} is $\e^{\beta h_1/2} \left(1 + \e^{-\beta h_1}\right)$. This equals $\Zf$, up to a finite prefactor. The counterterm that needs to be added to $\~S$ in order to compute $\Zf$ is thus $\frac12\beta h_1 + 2n\_S \log \d\tau$; this adjusts both the finite prefactor and the cutoff-dependent (nonuniversal) prefactor. As before, the finite part of this counterterm is solely determined by matching with the known exact answer.

The smooth action \eqref{def S tilde ferm} is ubiquitous in the literature. The smoothing step \eqref{temporal smoothing ferm} is never emphasized openly. That this omission is not quite harmless was demonstrated by the discussion of dilatations at the end of Subsection \ref{subsec eval path int}. The same discussion would apply here; it is not necessary to repeat it. Instead, the next Subsection will provide a different example in which temporal smoothness must be taken into account.

\subsection{Smoothing, two-point functions, and contact terms}

For simplicity, in this Subsection the Hamiltonian is simply
\bel{
  H = h \psi\+ \psi.
}
Consider the two-point correlation function
\bel{\label{def G}
  G(\tau) \equiv \avg{\psi\+ \psi_\tau}_\beta \equiv \Zf^{-1} \Tr\left[\psi\+ \e^{-\tau H} \psi\, \e^{-(\beta - \tau) H} \right].
}
This correlator is easily evaluated at any $\tau$, as the commutation relations imply
\bel{
  \psi_\tau \equiv \e^{-\tau H} \psi\, \e^{\tau H} = \e^{\tau h} \psi, \quad \psi\+_\tau \equiv \e^{-\tau H} \psi\+ \e^{\tau H} = \e^{-\tau h} \psi\+
}
and hence
\bel{\label{correl ferm 0 tau}
  G(\tau) = \Zf^{-1} \e^{-(\beta - \tau)h} = \frac{\e^{\tau h}}{1 + \e^{\beta h}}.
}
Note that in this QM problem there is nothing remotely singular happening as $\tau \rar 0$.

A path integral expression for $G(\tau)$ can be found by using the trick \eqref{Berezin multi trick} for a product of $N_0 + 2$ operators: two fermion operators and $N_0$ copies of $\e^{-\d\tau H}$. For the path integral to be nice, the operators in the correlation function must be time-ordered, which in this case amounts to requiring $\tau \geq \d\tau$. Then a conventional calculation shows that
\bel{
  G(\tau) = \Zf^{-1} \int [\d\eta \d\bar\eta] \, \bar\eta_{\d\tau} \eta_\tau \, \e^{-S}.
}
The action used here is given by \eqref{def S ferm}.

The frequency space technique (temporal smoothing) of Subsection \ref{subsec eval path int} is needed in order to actually do something with this path integral. As before, this entails keeping only low-frequency variables in the action and simply discarding the rest. However, there is now an extra fly in the ointment: the product $\bar \eta_{\d\tau} \eta_\tau$. These variables contain high-frequency modes that must be discarded, which amounts to replacing $\eta_\tau$ and $\bar\eta_{\d\tau}$ with their temporally smeared versions $\eta(\tau)$ and $\bar\eta(\d\tau) \approx \bar\eta(0)$ from \eqref{temporal smoothing ferm}. Doing so has two consequences:
\begin{enumerate}
  \item The path integral can be schematically expressed as $\sum_{\tau', \tau''} f(\tau', \d\tau) f(\tau, \tau'') \int \bar\eta_{\tau'} \eta_{\tau''} \e^{-S}$ where $f$ is a smearing function akin to the one used in \eqref{def varphi}. If $\tau$ and $\d\tau$ are sufficiently close to each other, the sum will involve path integrals with both time orderings ($\tau' < \tau''$ and $\tau' > \tau''$). In this regime the smooth path integral cannot correspond to a smearing of the simple correlation function $\avg{\psi\+ \psi_\tau}_\beta$ given by \eqref{correl ferm 0 tau}. The discontinuity due to smearing a time-ordered product must manifest itself as some kind of transition at small $\tau$.
  \item The partition function was given by a product over frequencies, so temporal smoothing resulted in an answer $\~\Zf$ that differed from the exact result $\Zf$ by an overall multiplicative prefactor that could be viewed as a counterterm in the smooth action $\~S$. The correlation function obtained by dropping high-frequency modes of Grassmann variables outside of the exponent $\e^{-S}$ may differ from the exact answer \eqref{correl ferm 0 tau} by an \emph{additive} term that cannot be cancelled out by a counterterm in $\~S$. This is a toy example that illustrates the origin of \emph{contact terms} in correlation functions \cite{Closset:2012vp}. In this case the finite part of the contact term will turn out to vanish.
\end{enumerate}

To illustrate these points explicitly, consider the smooth proxy for the correlator \eqref{def G}
\bel{\label{def G tilde}
  \~G(\tau) \equiv \~\Zf^{-1} \int [\d\eta \d\bar\eta] \, \bar\eta(0) \eta(\tau) \, \e^{-\~S},
}
for $\tau \geq 0$. The smooth action is
\bel{
  \~S = \sum_{n = -n\_S}^{n\_S - 1} \d\tau (-\i \omega_n + h)\, \bar\eta_n \eta_n = \sum_{\tau' = \d\tau}^{\beta} \d\tau\, \big[ -\bar\eta(\tau') \del_\tau \eta(\tau') + h\, \bar\eta(\tau') \eta(\tau') \big].
}
This path integral can be evaluated in frequency space without fancy regularization methods:
\algns{
  \~G(\tau)
  &= \frac1{\~\Zf N_0} \sum_{m,\, l = -n\_S}^{n\_S - 1} \e^{\i \, \omega_l \tau} \int [\d\eta\d\bar\eta]\, \bar\eta_m \eta_l \, \e^{\sum_{n = -n\_S}^{n\_S - 1} \d\tau (\i\omega_n - h) \bar\eta_n \eta_n}\\
  &=  \sum_{m = -n\_S}^{n\_S - 1} \frac{\e^{\i \, \omega_m \tau}}{\beta(\i \omega_m - h)}
  = \frac1\pi \int_{\pi/\beta}^{\omega\_S} \d\omega  \frac{\omega \sin \omega \tau - h \cos \omega \tau}{\omega^2 + h^2},
}
where $\omega\_S \equiv \frac{2\pi}\beta(n\_S + \frac12)$. The integral is simplest to do order-by-order in $\tau$. To first order,
\algns{\label{correl ferm 0 tau approx}
  \~G_1(\tau)
  &= \frac1\pi \left[\tau \left(\omega\_S - \frac\pi\beta \right)- (1 + \tau h) \left(\arctan\frac{\omega\_S}h - \arctan \frac\pi{\beta h} \right)\right] \approx  2n\_S \frac{\tau}\beta - C \e^{\tau h},
}
where $C \equiv \frac1\pi \left(\arctan\frac{\omega\_S}h - \arctan \frac\pi{\beta h} \right)$  lies between $-\frac12$ and $\frac12$. Including higher powers of $\tau$ precisely preserves the ``universal'' answer $-C \e^{\tau h}$ while changing the $n\_S$-dependent terms, which are now additive and not just multiplicative (consequence 2 above). 

Expanding in $\tau$ is justified if
\bel{\label{G1 validity}
  \tau \ll \frac1{\omega\_S} \approx \frac{\beta}{2\pi n\_S}.
}
This is also the regime in which the path integral has the field insertions $\bar\eta_{\d\tau}$ and $\eta_\tau$ at less than the smearing length ($\beta/2n\_S$) from each other. Thus $\~G_1(\tau)$ contains the information on the small-$\tau$ transition of $\~G(\tau)$ that was above advertised as consequence 1.

It is possible to examine this transition in detail. To start, note that the nonuniversal term in \eqref{correl ferm 0 tau approx} is precisely $\tau$ divided by the smearing length, and so in the regime of interest \eqref{G1 validity} this term is much smaller than unity. If the energy scale of the Hamiltonian, $h$, satisfies the reasonable condition
\bel{
  h \ll \omega\_S, \quad \trm{or} \quad 2n\_S \gg \beta h,
}
then \eqref{G1 validity} also implies $h\tau \ll 1$ and the correlator is simply
\bel{
  \~G(\tau) 
  \approx \frac1\pi \arctan\frac\pi{\beta h} + 2n\_S\frac{\tau}{\beta} - \frac12.
}
If $\beta h \sim 1$, this correlator is thus dominated by a $\tau$-independent constant, in agreement with the behavior of $G(\tau)$ as $\tau \rar 0$. However, at sufficiently low temperatures, such that
\bel{
  \beta h \gg 1,
}
the correlator is
\bel{
  \~G(\tau) \approx 2n\_S \frac \tau \beta - \frac{\beta h}{\pi^2}.
}
The advertised transition happens when the two terms become comparable, at
\bel{
  \tau_\star \approx \frac{\beta^2 h}{2\pi^2 n\_S}.
}
Around the point $\tau_\star$ the low-temperature correlator ceases to be dominated by the universal term (in this case $-\beta h/\pi^2$) and becomes sensitive to the cutoff $\omega\_S$. From the Hamiltonian point of view, this is unphysical, as $\omega\_S$ has no meaning in the canonical formalism. Indeed, the exact answer \eqref{correl ferm 0 tau} is not dominated by the linear behavior $\sim \omega\_S \tau$ at small times.

This analysis collapses if the smoothness scale is much smaller than the scale of the Hamiltonian, i.e.\ if
\bel{
  h \gg \omega\_S, \quad \trm{or} \quad 2n\_S \ll \beta h.
}
The correlator is then
\bel{
  \~G(\tau) \approx \frac{\omega\_S}\pi \left( \tau - \frac1h\right).
}
The point $\tau_\star \sim 1/h$ is harder to interpret as a physical crossover, as $\~G(\tau)$ never loses its explicit dependence on $\omega\_S$. This is not surprising: if $\omega\_S$ is not the largest scale in the smooth theory, then the action $\~S$ can never forget about its existence and return a universal answer.

Throughout this Subsection, terms of order $O(n\_S/N_0)$ were immediately dropped. It is possible to include them into the analysis and to find that the analogue of the roughening transition is present whenever $\omega\_S \tau_\star \ll n\_S/N_0$. This will not be explored here.
\newpage

\section{Supersymmetric QM} \label{sec SQM}

The final topic of this paper concerns the construction of supersymmetric (SUSY) continuum theories from lattice ones. This Section will present an eclectic mix of ideas, laying out the groundwork rather carefully, in an elementary yet unusually general way. The starting point will be a definition of SUSY applicable to any quantum theory, either lattice or continuum (Subsection \ref{subsec def susy}). The conventional QM representation of the SUSY structure \cite{Gendenshtein:1986ub} --- a coupling of a bosonic quantum particle and a fermion (two-state) system --- will be presented in Subsection \ref{subsec susy theory}. An amusing part of this analysis will be the construction of the ``minimal SUSY model,'' a theory with a four-dimensional Hilbert space and without a Lagrangian formulation at low temperatures. Finally, in Subsection \ref{subsec susy sho} the SUSY SHO will be tackled in an analysis that parallels that of Section \ref{sec sho}, and it will be shown how this familiar SUSY cQM emerges from the lattice one upon taming. All the theories examined in this Section will exhibit ``fermion doubling'' --- a double degeneracy of all the states in the spectrum, including the ground state --- that is a very general consequence of requiring SUSY to hold in a finite system that comprises a fermionic degree of freedom.

\subsection{The bare bones of supersymmetry} \label{subsec def susy}

In this paper, a theory is called \emph{supersymmetric} if it possesses a nilpotent symmetry, i.e.~an operator $Q$ that commutes with the Hamiltonian and obeys $Q^2 = 0$.\footnote{It should always be clear from context whether $Q$ refers to a SUSY generator or to a position operator.} If such a $Q$ is Hermitian, it must equal zero. If $Q$ is not Hermitian, $Q\+$ must also be a symmetry of the theory, since $[H, Q]\+ = [Q\+, H] = 0$. Further, $Q\+$ must also be nilpotent. Thus, if a theory has one nilpotent SUSY generator, or \emph{supercharge}, it must have at least one more.

The two generators $Q$ and $Q\+$ do not necessarily obey any further relations. At this point they simply generate the free algebra with two elements, i.e.~a vector space spanned by operators of the form $Q Q\+ Q Q\+ Q\cdots$. All of these commute with $H$. A single supercharge can thus imply a huge symmetry structure.

The anticommutator $\{Q, Q\+\}$ must commute with both supercharges. This is easy to check by computing
\bel{
  Q \{Q, Q\+\} = Q Q\+ Q = \{Q, Q\+\} Q.
}
The same will hold for anticommutators of the form
\bel{\label{def H n}
  \{Q (Q\+ Q)^n, (Q\+ Q)^n Q\+ \}.
}
Any of these anticommutators would be a reasonable choice for the Hamiltonian.

In fact, it is typical to \emph{demand} that a SUSY theory have, up to a rescaling,
\bel{\label{def H susy canonical}
  H = \{Q, Q\+ \}.
}
This will be the canonical choice in this paper. Nevertheless, it is important to keep in mind that there exist Hamiltonians with nilpotent symmetries that do not fulfill this criterion. Depending on one's outlook, imposing \eqref{def H susy canonical} as part of a definition of SUSY may be a natural thing to do, but many traits associated to SUSY do not require this choice.

A complementary point of view is obtained by defining the Hermitian supercharges
\bel{\label{def Q 12}
  \Q_1 \equiv \frac{Q + Q\+}2, \quad \Q_2 \equiv \frac{Q - Q\+}{2\i}.
}
They satisfy
\bel{
  \{\Q_1, \Q_2\} = 0, \quad \Q_1^2 = \Q_2^2 = \frac14 \{Q, Q\+\}.
}
These are generalized Majorana operators: they anticommute and their squares must be equal to each other, but they are not forced to square to the identity.

Now it is possible to define SUSY theories with any number $N\_{SUSY}$ of Hermitian supercharges  $\Q_I$. The requirement is that these satisfy
\bel{\label{def QI,QJ}
  \{\Q_I, \Q_J\} = 0 \ \trm{for}\ I \neq J, \quad \trm{and} \quad \Q_1^2 = \ldots = \Q_{N\_{SUSY}}^2.
}
When $N\_{SUSY} = 2n$, pairing up the $\Q_I$'s and taking linear combinations of pairs returns the nilpotent anticommuting generators, e.g.\ by letting $Q_i = \Q_{2i - 1} + \i \Q_{2i} $ for $1 \leq i \leq n$.  If $N\_{SUSY} = 2n + 1$, one can imagine extending this set of $Q_i$'s and $Q_i\+$'s by a single Hermitian operator $\Q_0$ that anticommutes with each $Q_i$ and obeys
\bel{
  \Q_0^2 = \frac14 \{Q_i, Q_i\+\}, \quad 1 \leq i \leq n.
}

It is interesting to consider the special case $N\_{SUSY} = 1$. Here there is only a Hermitian operator $\Q_0$ that does not need to satisfy any further restrictions. This is an ordinary symmetry. For instance, $\Q_0$ can be the fermion number $N\_F$ in a fermion QFT. Thus what is called SUSY in this paper can be understood as a generalization of ordinary symmetries to the case $N\_{SUSY} > 1$.

The case $N\_{SUSY} = 1$ further illuminates the implications of the optional condition \eqref{def H susy canonical}. This choice here amounts to demanding that $H = 4 \Q_0^2$, which is of course typically not required of a generic symmetry generator. Moreover, demanding \eqref{def H susy canonical} leads to the slightly absurd situation of saying that $H$ has a certain symmetry while e.g.~$H + c\1$ does not.

If the Hamiltonian is given by \eqref{def H susy canonical}, SUSY has two main dynamical consequences:
\begin{enumerate}
  \item If $\dim \H$ is even and $N\_{SUSY} \geq 2$, every energy level is (at least) doubly degenerate.
  \item All energies are nonnegative.
\end{enumerate}

To prove the first claim, take any two Hermitian SUSY generators $\Q_{1/2}$ and diagonalize one of them, say $\Q_1$, together with the Hamiltonian. Now the action of $\Q_2$ on any eigenstate $\qvec\psi$ of $\Q_1$ must give another eigenstate of $\Q_1$ with the opposite eigenvalue, since
\bel{
  \Q_1 \Q_2 \qvec \psi = - \Q_2 \Q_1 \qvec \psi = - \lambda \Q_2  \qvec\psi.
}
Thus $\Q_2 \qvec\psi$ must be orthogonal to $\qvec\psi$ if $\lambda \neq 0$. Since $[\Q_2, H] = 0$, the energy of $\Q_2 \qvec \psi$ and $\qvec \psi$ must be the same, so all eigenstates of $\Q_1$ with eigenvalues $\lambda \neq 0$ come in degenerate pairs. If the Hilbert space is even-dimensional, the number of states with $\lambda = 0$ must also be even. By \eqref{def H susy canonical} these must have zero energy, and the claim is established.

The second property is trivial, as $H = 4\Q_I^2$ directly implies that the Hamiltonian is nonnegative-definite. A more elementary proof is still useful, however, so here it is. If $\qvec \psi$ is a state with energy $E$, let $\qvec{\psi_I} \equiv \Q_I \qvec \psi$ for every $I$, and note that
\bel{
  E = \qmat \psi H \psi = 4\qmat \psi {\Q_I^2} \psi =  4\qprod{\psi_I}{\psi_I} \geq 0,
}
with equality achieved iff $\Q_I \qvec \psi = 0$. This completes the proof and makes clear that zero-energy (i.e.\ ground) states must be annihilated by every single supercharge $\Q_I$. In particular, in a smooth subspace it is easier to find the null space of $\Q_I$ than of $H \propto \Q_I^2$.

Both dynamical consequences hold for more general SUSY Hamiltonians \eqref{def H n}. Their study is left for the future.


\subsection{Two standard representations} \label{subsec susy theory}

The discussion so far has assumed nothing about the structure of the Hilbert space on which the SUSY generators act. Here there exists one choice that is so canonical as to be synonymous with SUSY: taking the theory to be a coupling of an arbitrary quantum theory (``bosons'') and a number of two-state theories (``fermions''). For $N\_{SUSY} = 2$, let
\bel{\label{B f convention}
  Q \equiv B\+ f, \quad Q\+ \equiv B f\+,
}
where $f\+$ and $f$ are, respectively, raising and lowering operators in a single fermionic theory, obeying $\{f, f\+\} = 0$ and $f^2 = (f\+)^2 = 0$, and $B$ is an arbitrary operator in the ``bosonic'' theory, e.g.\ the clock model. Let $n\_F \equiv f\+ f$ be the fermion number.

A simple Hamiltonian for which $Q$ and $Q\+$ are symmetries is found by using \eqref{def H susy canonical},
\bel{\label{H B f}
  H = \{Q, Q\+\} 
  = n\_F [B, B\+] + B\+ B.
}
Using Pauli matrices, $n\_F = \frac12(\1 - \sigma^z)$, gives the pleasant form
\bel{
  H = \frac12 \sigma^z [B\+, B] + \frac12 \{B\+, B\}.
}
In the special cases where $B$ is Hermitian or unitary, the fermion decouples from the boson, and the Hamiltonian is simply $H = B\+ B$. This is a trivial implementation of SUSY.

The \emph{minimal SUSY model} has another two-state system as its bosonic sector. The total Hilbert space is four-dimensional. Choose $B$ and $B\+$ to be the ladder operators analogous to $f$ and $f\+$, with the boson number $n\_B \equiv B\+ B$.\footnote{\, ``Bosonic'' and ``fermionic'' operators are here assumed to commute. If they were chosen to anticommute, so that e.g.\ $Bf = -fB$, the model would simply be a theory of two interacting fermions.  This theory would still exhibit SUSY. However, all formul\ae\ from \eqref{H B f} onwards must be altered in this case, as they were derived under the assumption that $B$ and $f$ commute.} Then the Hamiltonian \eqref{H B f} is
\bel{\label{def H min susy}
  H = n\_B + n\_F - 2 n\_B n\_F.
}
The spectrum of this Hamiltonian is $\{0, 0, 1, 1\}$, with zero energies corresponding to states with $n\_B = n\_F$. As expected, all states are doubly degenerate and have nonnegative energies.

The difficulty of recording this theory in path integral form is worth noting. By using $\e^{2\beta n\_B n\_F} = \1 + (\e^{2\beta} - 1) n\_B n\_F$, the partition function $\Zf = 2(1 + \e^{-\beta})$ can be expressed as
\bel{
  \Zf = \int [\d\chi \d\bar\chi \d\eta \d\bar\eta]\, \left[1 + (\e^{2\beta} - 1) \bar\eta_{\d\tau}\eta_{\d\tau} \bar\chi_{\d\tau} \chi_{\d\tau} \right] \e^{-S_0[\chi, \bar \chi] - S_0[\eta, \bar \eta]} ,
}
where
\bel{
  S_0[\eta, \bar \eta] \equiv \sum_{\tau = \d\tau}^\beta \d\tau \left(- \bar \eta_\tau (\del \eta)_{\tau - \d\tau} + \bar\eta_\tau \eta_\tau \right).
}
Translation invariance along the thermal circle means that
\bel{
  \Zf = \int [\d\chi \d\bar\chi \d\eta \d\bar\eta]\, \left[1 + \frac{\e^{2\beta} - 1}\beta \sum_{\tau = \d\tau}^\beta \d\tau \, \bar\eta_{\tau}\eta_{\tau} \bar\chi_{\tau} \chi_{\tau} \right] \e^{-S_0[\chi, \bar \chi] - S_0[\eta, \bar \eta]}.
}
At high temperatures ($\beta \ll 1$), the term in brackets can be written as the exponential of a four-fermion interaction. At low temperatures, there is no justification for re-exponentiation. In this sense the minimal SUSY model is a nonlagrangian theory. This is ultimately because the interaction strength in \eqref{def H min susy} cannot be decreased without breaking SUSY.

Going back to this Subsection's main line of development, it remains to touch upon the \emph{other} standard representation of SUSY algebras, used when discussing ``infinitesimal'' SUSY transformations. To define these, it is customary to introduce a Grassmann parameter $\eps$ that parameterizes a SUSY transformation. This is typically viewed as a formal extension of the scalar field over which QM operator algebras are defined, with $\C$ being replaced by a unital Grassmann algebra with a single Grassmann generator $\eps$. The object $\eps$ is then assumed to anticommute with all fermion-odd operators in the representation \eqref{B f convention}. There is nothing particularly infinitesimal about this parameter: its Grassmann nature simply allows any function $f(\eps)$ to be expressed as a linear polynomial in $\eps$.

There is an alternative but unfamiliar way of viewing this procedure. Its advantage is that the rules of QM stay intact --- the operator algebras remain defined over $\C$, and there are no c-numbers that fail to commute with operators. The idea is to add an ancillary fermion to the Hilbert space, with $\eps$ being a Majorana operator acting on this extra degree of freedom. The supercharges are then represented as matrices on this enlarged Hilbert space,
\bel{\label{def Q eps}
  Q_\eps \equiv Q \eps = B\+ f \eps , \quad Q\+_\eps = \eps B f\+ .
}
With $\eps$ normalized such that $\eps^2 = \1$, the candidate Hamiltonian $\{Q_\eps, Q\+_\eps\}$ is the same as the old one, \eqref{def H susy canonical}.

The utility of this representation is more evident after defining the SUSY variations
\bel{
  \delta_\eps \O \equiv [Q_\eps, \O], \quad \delta_\eps\+ \O \equiv [Q_\eps\+, \O].
}
These nilpotent operations act on the $B$ and $f$ operators in a familiar fashion,
\gathl{\label{SUSY algebra}
  \delta_\eps f\+ =  -\eps B\+ \{f, f\+\} = -\eps B\+, \quad \delta_\eps f = 0, \quad \delta_\eps\+ f\+ = 0, \quad \delta_\eps\+ f = \eps B \{f, f\+\} = \eps B, \\
  \delta_\eps B\+ = 0, \quad \delta_\eps B = [B, B\+]  \eps f \approx \eps f, \quad \delta_\eps\+ B\+ = [B, B\+] \eps f\+ \approx \eps f\+, \quad \delta_\eps\+ B = 0.
}
The approximations hold only in specific circumstances, e.g.\ if the bosonic system is a clock model that can be tamed, as described in Subsection \ref{subsec taming}. In this case it can be said that SUSY variations of bosons are fermions, and vice versa. The lesson of this derivation is that a ``superalgebra'' structure, schematically given by $\delta f \sim \eps B$ and $\delta B \sim \eps f$, is by no means germane to a finite theory with SUSY.

The representation \eqref{def Q eps} is naturally connected to the point of view advocated in Subsection \ref{subsec Berezin}, that a Grassmann variable can always be viewed as an operator in an auxiliary fermion system. Introducing a ``Grassmann-valued scalar,'' as is customary when discussing superalgebras or supergroups in a SUSY theory, is really equivalent to taking a graded direct product of this theory with the auxiliary system on which the Grassmann acts.
\newpage

\subsection{SUSY harmonic oscillator} \label{subsec susy sho}


In this Subsection, supercharges will be represented in the conventional form \eqref{B f convention}, with the bosonic system being a clock model of the kind studied in Sections \ref{sec clock} and \ref{sec sho}. The bosonic parts of the supercharges will be
\bel{\label{def B QM}
  B \equiv \i P + W(Q),
}
where $P$ and $Q$ are momentum and position operators defined in \eqref{def P} and \eqref{def Q}. The function $W(Q)$ will be called a \emph{superpotential}.\footnote{Sometimes one writes $B = \i P + W'(Q)$ and calls this $W(Q)$ the superpotential.}

For every superpotential $W(Q)$, it is possible to generate a sequence of SUSY Hamiltonians given by \eqref{def H n}. The simplest one, given by \eqref{H B f}, takes the form
\bel{\label{def H PW}
  H = P^2 + W^2(Q) + \i\, (2 n\_F - \1)  \big[P, W(Q)\big] .
}
Choosing $W(Q) = \omega_0 Q$ gives the Hamiltonian of the SUSY SHO. This theory can be viewed as a fermion (particle with two internal states) moving on a circle in the presence of a background potential, with SUSY relating the potentials felt by different fermion components.

The operators $P$ and $Q$ do not obey the canonical commutation relations. Just as with the ordinary SHO, it is first necessary to establish that this Hamiltonian has a low-energy eigenspace of tame states. Assuming this is true for the moment, it then makes sense to tame the Hamiltonian and get the more conventional form
\bel{\label{def H sho susy}
  H\_T \approx P^2 + \omega^2_0 Q^2 + \omega_0 \, (2 n\_F - \1).
}
After taming, the bosonic and fermionic systems decouple, but SUSY ensures that exciting the fermion costs the same amount of energy ($2\omega_0$) as exciting the SHO by one level. This gives rise to the expected double degeneracy in the spectrum --- the single exception being the ground state, as there is only one tame state with zero energy. This is incompatible with the statement that all states in the theory \eqref{def H PW} must be at least doubly degenerate, as follows from the general considerations of Subsection \ref{subsec def susy}.\footnote{Note that any theory represented as a product of a bosonic and a fermionic system has an even-dimensional Hilbert space, which is a necessary condition for the double degeneracy to be present.} Thus, without ever numerically diagonalizing the Hamiltonian, it becomes clear that the tame eigenspace \emph{cannot} be the only low-energy eigenspace in the theory.

The situation is analogous to the one encountered when comparing the SHO Hamiltonians \eqref{def H sho clock} and \eqref{def H sho PQ}. The low-energy eigenspace in the former case is tame; in the latter case, $H_\omega \sim P^2 + \omega^2 Q^2$ has three other low-energy eigenspaces, with eigenfunctions differing from tame ones by a $(-1)^\phi$ modulation or by a shift along the target by $\Delta\phi = \pi$.

The fourfold degeneracy in the nonsupersymmetric theory \eqref{def H sho PQ} can be associated to the ``spontaneous breaking'' of the $\Z_2 \times \Z_2$ symmetry generated by transformations $Z \mapsto -Z$ and $X \mapsto -X$. (This simply means that each of the ground states has a different charge under the $\Z_2 \times \Z_2$.) Coupling this theory to a fermion to get the SUSY Hamiltonian \eqref{def H PW} explicitly breaks this $\Z_2 \times \Z_2$ symmetry down to a single $\Z_2$ that acts as $(Z, X) \mapsto -(Z,X)$. This symmetry is generated by $(ZX)^{K/2}$. The fermion number $n\_F$ is also conserved, however, so the theory \eqref{def H PW} is again found to have a $\Z_2 \times \Z_2$ symmetry. 

Numerically diagonalizing the Hamiltonian \eqref{def H PW} reveals that it has a fourfold ground state degeneracy that corresponds to the ``spontaneous breaking'' of this new $\Z_2 \times \Z_2$, in the same sense as in the previous paragraph. At $n\_F = 0$, its ground state eigenfunctions are a smooth Gaussian localized around $\phi = 0$ and its image under $(ZX)^{K/2}$, a $(-1)^\phi$-modulated Gaussian localized around $\phi = \pi$. At $n\_F = 1$, the ground state wavefunctions are a smooth Gaussian around $\phi = \pi$ and a $(-1)^\phi$-modulated Gaussian around $\phi = 0$. These can be organized into eigenspaces of the $\Z_2 \times \Z_2$ generators. In practice, one ignores the $\Z_2 \times \Z_2$ symmetry and isolates the tame subspace of the bosonic sector; only one of these four ground states belongs to this subspace, and it has $n\_F = 0$. This subspace is \emph{not} an eigenspace of the whole $\Z_2 \times \Z_2$, but it is an eigenspace of the fermionic $\Z_2$ generated by $(-1)^{n\_F}$.

This last observation makes it natural to define the Witten index, which is essentially a grand canonical partition function with a nonzero chemical potential for fermion parity,
\bel{\label{def Witten index}
  \fr W \equiv \Tr\left[ (-1)^{n\_F} \e^{-\beta H}\right].
}
The celebrated property of the Witten index is that it is invariant under most deformations of the Hamiltonian; when the theory is given by \eqref{def H PW}, $\fr W$ only depends on certain global properties of the superpotential \cite{Witten:1982df}. However, in the context of this paper, this claim must be parsed more carefully. The index defined by \eqref{def Witten index} is necessarily zero in \emph{any} theory where SUSY is represented via \eqref{B f convention}. This follows from the double degeneracy proven in Subsection \ref{subsec def susy}. It is only the index over tame states,
\bel{
  \fr W\_T \equiv \Tr_{\H\_T} \left[ (-1)^{n\_F} \e^{-\beta H}\right],
}
that has any nontrivial properties. (Compare this to the smooth partition function \eqref{def ZfS}.)


Ground state properties in the tame sector can be inferred from SUSY without ever numerically diagonalizing the Hamiltonian. (Of course, this requires assuming that such a state exists in the first place, and working self-consistently from there on out.) Requiring that both $B\+ f$ and $Bf\+$ annihilate the ground state and assuming that $B$ and $B\+$ are SHO ladder operators in a tame space implies that the ground state has quantum numbers $n\_F = n\_B = 0$. This works for other superpotentials, and it is one of the main draws of SUSY.

It is less widely appreciated that this commonly touted power of SUSY is curbed by the need to tune the couplings in the superpotential so that the microscopic Hamiltonian \eqref{def H PW} actually has tame eigenstates (see Fig.\ \ref{fig support}). Numerics shows that, for $W(Q) = \omega_0 Q$, the coupling must satisfy
\bel{
  \omega_0 \approx \frac K{2\pi} = \frac1{\d\phi},
}
in the same sense that $\gamma \approx 1$ was required in Section \ref{sec sho}. It is thus natural to define
\bel{
  \omega \equiv \omega_0 \d\phi
}
as the $O(1)$ coupling in the conveniently rescaled cQM Hamiltonian
\bel{
  H\_c \equiv \frac12(\d\phi)^2 H\_T \approx  \frac12(\d\phi)^2 P^2 + \frac12 \omega^2 Q^2 + \frac12\d\phi\, \omega \, (2 n\_F - \1).
}
The factors of $\d\phi$ that appear throughout this Hamiltonian perform the same roles that $\hbar$ does in much of the literature: they control how close the theory is to the ``critical'' point at which the continuum description is valid. Importantly, however, $\d\phi = 2\pi/K$ has a clear interpretation as a dimensionless number that cannot be freely ``set to unity,'' as one may be tempted to do with $\hbar$ after too much exposure to perturbative QFT.

The upshot of this is that the continuum coupling $\omega$ cannot be taken to an extreme in either direction: if
\bel{
  \omega \lesssim \d\phi \quad \trm{or} \quad \omega \gtrsim \frac1{\d\phi},
}
the assumption of tameness becomes inconsistent with the existence of the microscopic Hamiltonian \eqref{def H PW}, and the canonical commutation relations in the bosonic sector (and hence also the continuum Hamiltonian) start receiving large corrections. As before, these are weak bounds that merely serve to prove a point --- that SUSY cQMs cannot exist at arbitrary couplings. Finding stronger results remains an open question.

To summarize, this Section has stressed the following unusual points:
\begin{enumerate}
  \item It is possible to define SUSY theories with finite Hilbert spaces.
  \item SUSY Hamiltonians need not be simple anticommutators of two supercharges.
  \item The Witten index of any lattice SUSY theory represented as a coupling of fermions and bosons is necessarily zero. All states must be at least doubly degenerate.
  \item Many familiar properties of SUSY appear only after restricting to a tame subspace. However, doing so may mean missing out on entire sectors of low-energy states. The restriction itself may be inconsistent at extreme values of the coupling.
\end{enumerate}

\section{Summary}

This paper has presented a broad array of examples that showcase the utility of the smoothing perspective in QM. It may be helpful to recap the highlights and stress their implications for the lattice-continuum correspondence:
\begin{enumerate}
  \item The operator algebras of four ubiquitous cQM theories --- the free particle on $\R$ and on $S^1$, and the simple and supersymmetric harmonic oscillators --- can be realized to arbitrary precision via controlled reductions (smoothings and/or tamings) of a sufficiently large  but finite clock algebra \eqref{def Z X rels}.
  \item An analogous reduction can be expressed in the path integral language by restricting the set of states inserted at each time step. When the parameters of this reduction are chosen judiciously (relative to the original Hilbert space dimension and the size of the time step), the resulting actions take on the familiar quadratic forms, and moreover the path integrals are good approximations to exact answers, as per eq.\ \eqref{alpha bound} and Subsection \ref{subsec bounds}. This analysis also leads to a lattice-based definition of sums over saddle points and of the distinction between compact and noncompact QM theories.
  \item To actually evaluate the path integrals, an uncontrolled procedure of temporal smoothing must be used, as explained in Subsection \ref{subsec eval path int}: the sum over Matsubara frequencies is manually restricted, and a universal answer is extracted after including the appropriate counterterms. This analysis leads to a purely lattice-based understanding of the origin of counterterms and contact terms in correlation functions, as well as to a precise description of the emergence of spacetime symmetries.
  \item Fermionic QM theories do not admit a reduction to a continuum theory in the canonical formalism, but their path integrals can be evaluated after performing the same uncontrolled temporal smoothing that was employed in the bosonic case. It is after these manipulations that bosonic and fermionic actions become closely analogous to each other.
  \item It is possible to discuss SUSY in a very general lattice setting, and a simple nonlagrangian theory with a finite Hilbert space, the ``minimal SUSY model,'' was given in eq.\ \eqref{def H min susy}. However, it is only after taming that the symmetry between bosons and fermions comes in sharp relief, with the emergence of the SUSY algebra structure \eqref{SUSY algebra}. This analysis has shown how such an algebra arises from a lattice model, and along the way it was also proven that the Witten index of any lattice theory with exact SUSY must be zero.
\end{enumerate}

Target space smoothing and taming are key ingredients in the lattice-continuum correspondence of higher-dimensional QFTs. A conventional scalar cQFT can be obtained from a lattice by taming the target space at each spatial point, followed by smoothing along spatial directions. The procedure of smoothing along spatial directions, first developed in the context of simple fermionic QFTs \cite{Radicevic:2019jfe, Radicevic:2019mle}, will be presented in much greater generality in the next part of this series.

There do remain questions regarding the lattice-continuum correspondence in QM that this paper has not addressed. For example, how does reparameterization invariance of the action precisely emerge from a finite system, as seen in the SYK model \cite{Kitaev:2015}?  Or, for instance, how does the hierarchy of scales needed to define a cQM appear within the (hyper)asymptotics of cQM perturbation theory \cite{Berry:1989, Basar:2013eka}?

\section*{Acknowledgments}

It is a pleasure to thank Nathan Benjamin, Matt Headrick, and Mithat \"Unsal for useful conversations. This work was completed with the support from the Simons Foundation through \emph{It from Qubit: Simons Collaboration on Quantum Fields, Gravity, and Information}, and from the Department of Energy Office of High-Energy Physics grant DE-SC0009987 and QuantISED grant DE-SC0020194. Part of this work was carried out at the Kavli Institute for Theoretical Physics, with support from the National Science Foundation under Grant No.\ NSF PHY-1748958.

\bibliographystyle{ssg}
\bibliography{Refs}

\begingroup\raggedright\begin{thebibliography}{10}

\bibitem{Wilson:1971bg}
K.~G. Wilson, ``{Renormalization group and critical phenomena. 1.
  Renormalization group and the Kadanoff scaling picture},'' {\em Phys. Rev.}
  {\bf B4} (1971) 3174--3183.

\bibitem{Cardy:1996xt}
J.~L. Cardy, {\em {Scaling and renormalization in statistical physics}}.
\newblock Cambridge University Press, 1996.

\bibitem{DiFrancesco:1997nk}
P.~Di~Francesco, P.~Mathieu, and D.~Senechal, {\em {Conformal Field Theory}}.
\newblock Graduate Texts in Contemporary Physics. Springer-Verlag, New York,
  1997.

\bibitem{Atiyah:1988}
M.~Atiyah, ``Topological quantum field theories,'' {\em Publications
  Mathématiques de l’Institut des Hautes Scientifiques} {\bf 68} (1988)
  175--186.

\bibitem{Haag:1996}
R.~Haag, {\em {Local quantum physics. Fields, particles, algebras}}.
\newblock Springer-Verlag, Berlin, 1996.

\bibitem{Segal:1988}
G.~Segal, ``The definition of conformal field theory,'' in {\em Topology,
  geometry and quantum field theory} (U.~Tillmann, ed.), vol.~308,
  pp.~421--577.
\newblock Cambridge University Press, 2004.

\bibitem{Pokrovsky:1968}
V.~Pokrovskii, ``Similarity hypothesis in the theory of phase transitions,''
  {\em Soviet Physics Uspekhi} {\bf 11} (1968), no.~1 66.

\bibitem{Polyakov:1970xd}
A.~M. Polyakov, ``{Conformal symmetry of critical fluctuations},'' {\em JETP
  Lett.} {\bf 12} (1970) 381--383.

\bibitem{Chen:2011pg}
X.~Chen, Z.-C. Gu, Z.-X. Liu, and X.-G. Wen, ``{Symmetry protected topological
  orders and the group cohomology of their symmetry group},'' {\em Phys. Rev.}
  {\bf B87} (2013), no.~15 155114, \href{https://arxiv.org/abs/1106.4772}{{\tt
  1106.4772}}.

\bibitem{tHooft:1979rat}
G.~'t~Hooft, ``{Naturalness, chiral symmetry, and spontaneous chiral symmetry
  breaking},'' {\em NATO Sci. Ser. B} {\bf 59} (1980) 135--157.

\bibitem{Wen:2013oza}
X.-G. Wen, ``{Classifying gauge anomalies through symmetry-protected trivial
  orders and classifying gravitational anomalies through topological orders},''
  {\em Phys. Rev.} {\bf D88} (2013), no.~4 045013,
  \href{https://arxiv.org/abs/1303.1803}{{\tt 1303.1803}}.

\bibitem{Nielsen:1980rz}
H.~B. Nielsen and M.~Ninomiya, ``{Absence of Neutrinos on a Lattice. 1. Proof
  by Homotopy Theory},'' {\em Nucl. Phys.} {\bf B185} (1981) 20.

\bibitem{Nielsen:1981xu}
H.~B. Nielsen and M.~Ninomiya, ``{Absence of Neutrinos on a Lattice. 2.
  Intuitive Topological Proof},'' {\em Nucl. Phys.} {\bf B193} (1981) 173--194.

\bibitem{Radicevic:2019jfe}
{\DJ}.~Radi{\v c}evi{\'c}, ``{Abelian Bosonization, OPEs, and the `String
  Scale' of Fermion Fields},'' \href{https://arxiv.org/abs/1912.01022}{{\tt
  1912.01022}}.

\bibitem{Radicevic:2019mle}
{\DJ}.~Radi{\v c}evi{\'c}, ``{The Lattice-Continuum Correspondence in the Ising
  Model},'' \href{https://arxiv.org/abs/1912.13462}{{\tt 1912.13462}}.

\bibitem{Radicevic:2D}
{\DJ}.~Radi{\v c}evi{\'c}, ``{The Ultraviolet Structure of Quantum Field
  Theories. Part 2: What is Quantum Field Theory?},''
  \href{https://arxiv.org/abs/2105.12147}{{\tt 2105.12147}}.

\bibitem{Radicevic:3D}
{\DJ}.~Radi{\v c}evi{\'c}, ``{The Ultraviolet Structure of Quantum Field
  Theories. Part 3: Gauge Theories},''
  \href{https://arxiv.org/abs/2105.12751}{{\tt 2105.12751}}.

\bibitem{Radicevic:4D}
{\DJ}.~Radi{\v c}evi{\'c}, ``The ultraviolet structure of quantum field
  theories. part 4: Colors and flavors.'' In preparation.

\bibitem{Witten:1982df}
E.~Witten, ``{Constraints on Supersymmetry Breaking},'' {\em Nucl. Phys. B}
  {\bf 202} (1982) 253.

\bibitem{Schwinger:1960}
J.~S. Schwinger, ``{Unitary Operator Bases},'' {\em Proc.\ Natl.\ Acad.\ Sci.\
  USA} {\bf 46} (1960) 570–579.

\bibitem{Gelfand:1964}
I.~M. Gel'fand and N.~J. Vilenkin, {\em Generalized Functions, vol. 4: Some
  Applications of Harmonic Analysis. Rigged Hilbert Spaces}.
\newblock Academic Press, New York, 1964.

\bibitem{Reed:1972}
M.~Reed and B.~Simon, {\em Methods of modern mathematical physics I: Functional
  analysis}.
\newblock Academic Press, New York, 1972.

\bibitem{Radicevic:2016kpf}
{\DJ}.~Radi{\v c}evi{\'c}, ``{Quantum Mechanics in the Infrared},''
  \href{https://arxiv.org/abs/1608.07275}{{\tt 1608.07275}}.

\bibitem{Balasubramanian:2018axm}
V.~Balasubramanian and O.~Parrikar, ``{Remarks on entanglement entropy in
  string theory},'' {\em Phys. Rev. D} {\bf 97} (2018), no.~6 066025,
  \href{https://arxiv.org/abs/1801.03517}{{\tt 1801.03517}}.

\bibitem{Mazenc:2019ety}
E.~A. Mazenc and D.~Ranard, ``{Target Space Entanglement Entropy},''
  \href{https://arxiv.org/abs/1910.07449}{{\tt 1910.07449}}.

\bibitem{Gupta:1993id}
K.~Gupta and S.~Rajeev, ``{Renormalization in quantum mechanics},'' {\em Phys.
  Rev. D} {\bf 48} (1993) 5940--5945,
  \href{https://arxiv.org/abs/hep-th/9305052}{{\tt hep-th/9305052}}.

\bibitem{Polonyi:1994pn}
J.~Polonyi, ``{Renormalization group in quantum mechanics},'' {\em Annals
  Phys.} {\bf 252} (1996) 300--328,
  \href{https://arxiv.org/abs/hep-th/9409004}{{\tt hep-th/9409004}}.

\bibitem{Ho:2017nyc}
W.~W. Ho and {\DJ}.~Radi\v{c}evi\'c, ``{The Ergodicity Landscape of Quantum
  Theories},'' {\em Int. J. Mod. Phys. A} {\bf 33} (2018), no.~04 1830004,
  \href{https://arxiv.org/abs/1701.08777}{{\tt 1701.08777}}.

\bibitem{Shenker:2013pqa}
S.~H. Shenker and D.~Stanford, ``{Black holes and the butterfly effect},'' {\em
  JHEP} {\bf 03} (2014) 067, \href{https://arxiv.org/abs/1306.0622}{{\tt
  1306.0622}}.

\bibitem{Kitaev:2015}
A.~Kitaev, ``A simple model of quantum holography.''
\newblock KITP strings seminar and Entanglement program (Feb 12, Apr 7, and May
  27, 2015),
  \href{http://online.kitp.ucsb.edu/online/entangled15/}{http://online.kitp.ucsb.edu/online/entangled15/}.

\bibitem{Cotler:2016fpe}
J.~S. Cotler, G.~Gur-Ari, M.~Hanada, J.~Polchinski, P.~Saad, S.~H. Shenker,
  D.~Stanford, A.~Streicher, and M.~Tezuka, ``{Black Holes and Random
  Matrices},'' {\em JHEP} {\bf 05} (2017) 118,
  \href{https://arxiv.org/abs/1611.04650}{{\tt 1611.04650}}. [Erratum: JHEP 09,
  002 (2018)].

\bibitem{Lin:2018bud}
J.~Lin and {\DJ}.~Radi{\v c}evi{\'c}, ``{Comments on defining entanglement
  entropy},'' {\em Nucl. Phys. B} {\bf 958} (2020) 115118,
  \href{https://arxiv.org/abs/1808.05939}{{\tt 1808.05939}}.

\bibitem{Dirac:1927dy}
P.~A. Dirac, ``{Quantum theory of emission and absorption of radiation},'' {\em
  Proc. Roy. Soc. Lond. A} {\bf 114} (1927) 243.

\bibitem{Carruthers:1968my}
P.~Carruthers and M.~M. Nieto, ``{Phase and angle variables in quantum
  mechanics},'' {\em Rev. Mod. Phys.} {\bf 40} (1968) 411--440.

\bibitem{Lynch:1995}
R.~{Lynch}, ``{The quantum phase problem: a critical review},'' {\em Phys.
  Repts} {\bf 256} (May, 1995) 367--436.

\bibitem{Susskind:1964zz}
L.~Susskind and J.~Glogower, ``{Quantum mechanical phase and time operator},''
  {\em Physics Physique Fizika} {\bf 1} (1964), no.~1 49--61.

\bibitem{Ghosh:2017pel}
S.~Ghosh and S.~Raju, ``{Loss of locality in gravitational correlators with a
  large number of insertions},'' {\em Phys. Rev. D} {\bf 96} (2017), no.~6
  066033, \href{https://arxiv.org/abs/1706.07424}{{\tt 1706.07424}}.

\bibitem{Kempf:1994su}
A.~Kempf, G.~Mangano, and R.~B. Mann, ``{Hilbert space representation of the
  minimal length uncertainty relation},'' {\em Phys. Rev. D} {\bf 52} (1995)
  1108--1118, \href{https://arxiv.org/abs/hep-th/9412167}{{\tt
  hep-th/9412167}}.

\bibitem{Chang:2011jj}
L.~N. Chang, Z.~Lewis, D.~Minic, and T.~Takeuchi, ``{On the Minimal Length
  Uncertainty Relation and the Foundations of String Theory},'' {\em Adv. High
  Energy Phys.} {\bf 2011} (2011) 493514,
  \href{https://arxiv.org/abs/1106.0068}{{\tt 1106.0068}}.

\bibitem{Jizba:2009qf}
P.~Jizba, H.~Kleinert, and F.~Scardigli, ``{Uncertainty Relation on World
  Crystal and its Applications to Micro Black Holes},'' {\em Phys. Rev. D} {\bf
  81} (2010) 084030, \href{https://arxiv.org/abs/0912.2253}{{\tt 0912.2253}}.

\bibitem{Radicevic:2018okd}
{\DJ}.~Radi{\v c}evi{\'c}, ``{Spin Structures and Exact Dualities in Low
  Dimensions},'' \href{https://arxiv.org/abs/1809.07757}{{\tt 1809.07757}}.

\bibitem{Polchinski:1998rq}
J.~Polchinski, {\em {String theory. Vol. 1: An introduction to the bosonic
  string}}.
\newblock Cambridge Monographs on Mathematical Physics. Cambridge University
  Press, 2007.

\bibitem{Itzykson:1980fz}
C.~Itzykson, M.~E. Peskin, and J.-B. Zuber, ``{Roughening of Wilson's
  Surface},'' {\em Phys. Lett. B} {\bf 95} (1980) 259--264.

\bibitem{Creutz:1999zy}
M.~Creutz, ``{Transfer matrices and lattice fermions at finite density},'' {\em
  Found. Phys.} {\bf 30} (2000) 487--492,
  \href{https://arxiv.org/abs/hep-lat/9905024}{{\tt hep-lat/9905024}}.

\bibitem{Headrick:2008ke}
M.~Headrick, ``{A solution manual for Polchinski's 'String Theory'},''
  \href{https://arxiv.org/abs/0812.4408}{{\tt 0812.4408}}.

\bibitem{Closset:2012vp}
C.~Closset, T.~T. Dumitrescu, G.~Festuccia, Z.~Komargodski, and N.~Seiberg,
  ``{Comments on Chern-Simons Contact Terms in Three Dimensions},'' {\em JHEP}
  {\bf 09} (2012) 091, \href{https://arxiv.org/abs/1206.5218}{{\tt 1206.5218}}.

\bibitem{Gendenshtein:1986ub}
L.~Gendenshtein and I.~Krive, ``{Supersymmetry in Quantum Mechanics},'' {\em
  Sov. Phys. Usp.} {\bf 28} (1985) 645--666.

\bibitem{Berry:1989}
M.~V. Berry, ``Some quantum-to-classical asymptotics,'' in {\em Les Houches
  Lecture Series LII} (M.-J. Giannoni, A.~Voros, and Z.~Zinn-Justin, eds.),
  pp.~251--304, North-Holland, Amsterdam.

\bibitem{Basar:2013eka}
G.~Basar, G.~V. Dunne, and M.~Unsal, ``{Resurgence theory, ghost-instantons,
  and analytic continuation of path integrals},'' {\em JHEP} {\bf 10} (2013)
  041, \href{https://arxiv.org/abs/1308.1108}{{\tt 1308.1108}}.

\end{thebibliography}\endgroup

\end{document}